\documentclass[useAMS,usenatbib]{mn2e}

\usepackage[latin1]{inputenc}
\usepackage{multirow}
\usepackage{graphicx}
\usepackage{multicol}
\usepackage{float}
\usepackage{color}
\usepackage{footnote}
\usepackage{amssymb}
\usepackage[figuresright]{rotating}
\usepackage{amsmath}
\usepackage{url}
\usepackage{rotating}
\usepackage{caption}
\newcommand{\apgt}{{\raise-.5ex\hbox{$\buildrel>\over\sim$}}}
\newcommand{\aplt}{{\raise-.5ex\hbox{$\buildrel<\over\sim$}}} 
\newcommand{\Mpc}{\rm\; Mpc}

\newcommand{\km}{\rm\; km}

\newcommand{\cm}{\rm\; cm}

\newcommand{\s}{\rm\; s}

\newcommand{\ks}{\rm\; ks}

\newcommand{\keV}{\rm\; keV}
\newcommand{\eV}{\rm\; eV}
\newcommand{\erg}{\rm\; erg}

\newcommand{\ergpcmsqps}{\hbox{$\erg\cm^{-2}\s^{-1}\,$}}

\newcommand{\ergps}{\hbox{$\erg\s^{-1}\,$}}

\newcommand{\kmps}{\hbox{$\km\s^{-1}\,$}}

\newcommand{\kmpspMpc}{\hbox{$\kmps\Mpc^{-1}\,$}}

\newcommand{\Lx}{\hbox{$\thinspace L_\mathrm{X}$}}

\newcommand{\Fx}{\hbox{$\thinspace F_\mathrm{X}$}}

\newcommand{\Omm}{\hbox{$\rm\thinspace \Omega_{m}$}}
\newcommand{\OmL}{\hbox{$\rm\thinspace \Omega_{\Lambda}$}}

\title[Rapid evolution of AGN feedback in BCGs]{The rapid evolution of AGN feedback in brightest cluster galaxies: switching from quasar-mode to radio-mode feedback}

\author[J. Hlavacek-Larrondo et al.]{J. Hlavacek-Larrondo$^{1,2,3}$\thanks{E-mail: juliehl@stanford.edu}\thanks{Einstein fellow}, A. C. Fabian$^{1}$, A. C. Edge$^{4}$, H. Ebeling$^{5}$, S. W. Allen$^{2,3,6}$, \newauthor{J. S. Sanders$^{1}$ and G. B. Taylor$^{7,8}$}\\
$^{1}$Institute of Astronomy, University of Cambridge, Madingley Road, Cambridge CB3 0HA\\
$^{2}$Kavli Institute for Particle Astrophysics and Cosmology, Stanford University, 382 Via Pueblo Mall, Stanford, CA 94305-4060, USA\\
$^{3}$Department of Physics, Stanford University, 452 Lomita Mall, Stanford, CA 94305-4085, USA\\
$^{4}$Institute of Computational Cosmology, Department of Physics, Durham University, Durham, DH1 3LE\\
$^{5}$Institute of Astronomy, University of Hawaii, 2680 Woodlawn Drive, Honolulu, HI 96822, USA\\
$^{6}$SLAC National Accelerator Laboratory, 2575 Sand Hill Road, Menlo Park, CA 94025, USA\\
$^{7}$Department of Physics and Astronomy, University of New-Mexico, Albuquerque, NM 87131, USA\\
$^{8}$National Radio Astronomy Observatory, Socorro, NM 87801, USA\\}
\begin{document}

\date{}

\pagerange{\pageref{firstpage}--\pageref{lastpage}} \pubyear{2012}

\maketitle

\begin{abstract}
We present an analysis of the $2-10\keV$ X-ray emission associated with the active galactic nuclei (AGNs) in brightest cluster galaxies (BCGs). Our sample consists of 32 BCGs that lie in highly X-ray luminous cluster of galaxies ($L_{\rm X-ray~}{\rm (0.1-2.4\keV)}>3\times10^{44}\ergps$) in which AGN-jetted outflows are creating and sustaining clear X-ray cavities. Our sample covers the redshift range $0<z<0.6$ and reveals strong evolution in the nuclear X-ray luminosities, such that the black holes in these systems have become on average at least 10 times fainter over the last 5 Gyrs. Mindful of potential selection effects, we propose two possible scenarios to explain our results: 1) either that the AGNs in BCGs with X-ray cavities are steadily becoming fainter, or more likely, 2) that the fraction of these BCGs with radiatively efficient nuclei is decreasing with time from roughly 60 per cent at $z\approx0.6$ to 30 per cent at $z\approx0.1$. Based on this strong evolution, we predict that a significant fraction of BCGs in $z\approx1$ clusters may host quasars at their centres, potentially complicating the search for such clusters at high redshift. In analogy with black-hole binaries and based on the observed Eddington ratios of our sources, we further propose that the evolving AGN population in BCGs with X-ray cavities may be transiting from a canonical low/hard state, analogous to that of X-ray binaries, to a quiescent state over the last 5 Gyrs. 
\end{abstract}


\section{Introduction}
The interplay between the accretion of material onto a supermassive black hole (SMBH) and the release of energy through radiation and outflows is known as active galactic nucleus (AGN) feedback. Some of the strongest cases of AGN feedback are seen in clusters of galaxies, where the central AGN hosted by the brightest cluster galaxy (BCG) is capable of driving large jetted outflows filled with relativistic plasma. As the jetted outflows propagate through the intracluster plasma, they push aside the hot X-ray emitting gas, creating cavities that are detectable as regions of reduced surface brightness in X-ray images. These X-ray cavities act as calorimeters and provide a unique opportunity to directly measure the work done by the AGN on the surrounding medium. 

X-ray cavities are therefore extremely useful tools for studying the details of AGN feedback. Studies at low redshifts have shown that these structures are predominantly found in cool-core clusters of galaxies \citep[detection rate $>90$ per cent;][]{Dun2006373,Fab2012}. Cool-core clusters have highly peaked X-ray surface brightness profiles, and central cooling times that are often shorter than the Hubble time. The hot X-ray gas at the centres of these clusters should therefore have had the time to cool, and large flows of cooling material, known as cooling flows, would naively be expected in the central regions of these objects \citep[see][]{Sar198658,Fab1984310,Fab199432}. However, both $Chandra$ and $XMM$ observations have shown that there is significantly less cooling material than expected from standard cooling-flow models \citep{Boh2001365,Tam2001365,Pet2001365,Pet2003590,Pet2006427,Mcn200745}. This is known as the cooling-flow problem, and feedback from the central AGN is thought to be the leading mechanism that prevents the hot X-ray gas from cooling, by inflating the X-ray cavities and propagating energy through shock and sounds waves \citep{Fab2006366,San2008390}. Studies of nearby clusters ($z<0.3$) have shown that the energetics of these X-ray cavities are indeed capable, on average, of preventing the hot X-ray gas from cooling \citep{Bir2004607,Bir2008686,Dun2005364,Dun2006373,Dun2008385,Nul2007,Dun2010404,Cav2010720,Don2010712,OSu2011735}, while simulations have begun to incorporate this kinetic mode of feedback in a cosmological context \citep[e.g.][]{Sij2007380,Dub2010409}.

\begin{figure}
\centering
\begin{minipage}[c]{0.99\linewidth}
\centering \includegraphics[width=\linewidth]{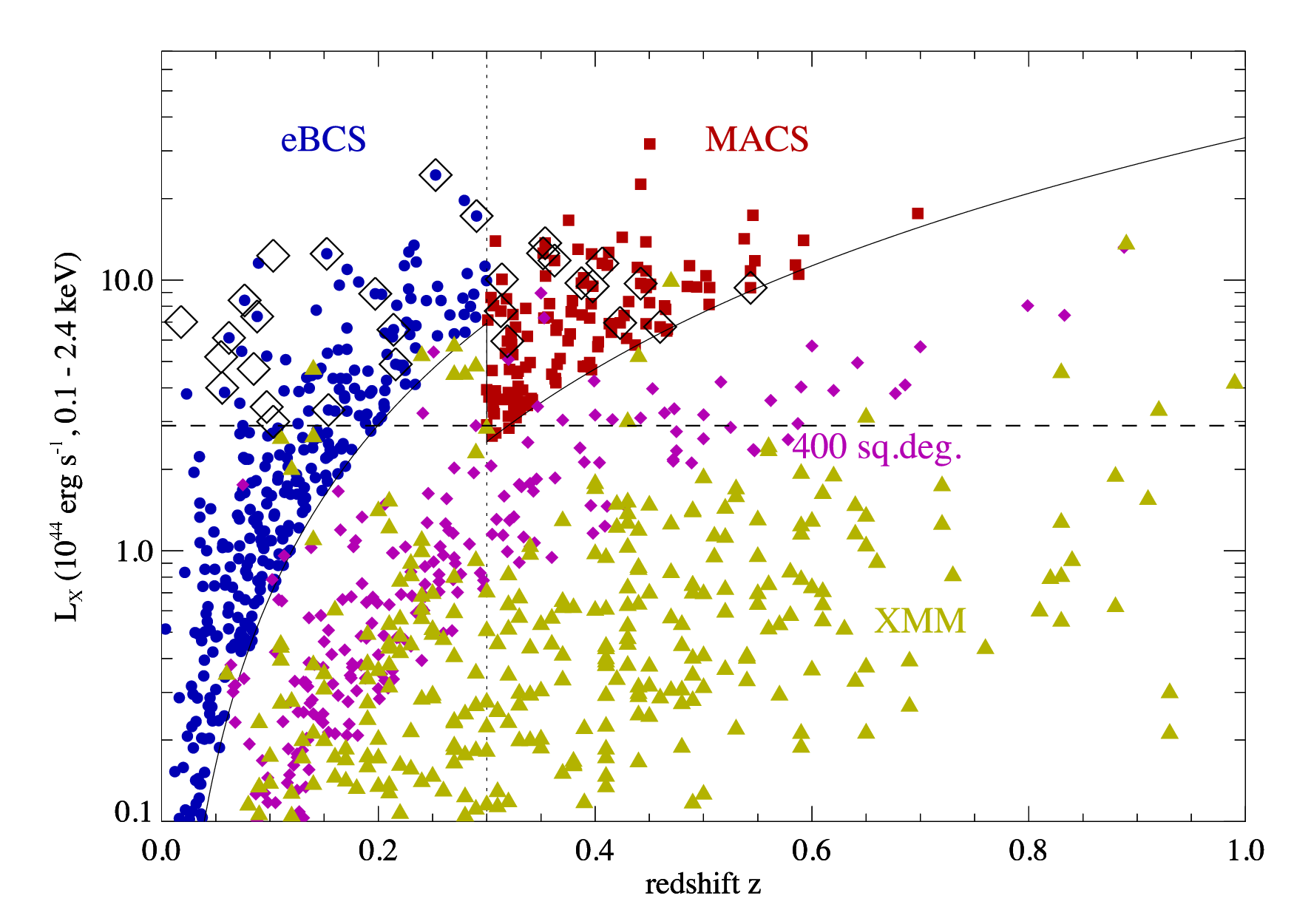}
\end{minipage}
\vspace{-0.in}
\caption[X-ray luminosity as a function of redshift for different cluster samples]{Cluster X-ray luminosity versus redshift. Shown are the extended Brightest Cluster Survey \citep[eBCS;][]{Ebe1998301,Ebe2000318}, the 400 deg$^2$ cluster survey \citep[][]{Bur2007172}, the $XMM-newton$ cluster survey \citep{Meh2012423} and the MACS survey which finds the luminous clusters at $0.3<z<0.7$. The black diamonds highlight the clusters in our sample and the horizontal dashed line shows our $>3\times10^{44}\ergps$ luminosity cut. The empty black diamonds show the clusters not part of eBCS. Here, we use the cluster luminosities from Table \ref{tab5_2} to illustrate them on the plot. Note that, we do not include 4C+55.16 and H1821+643 since they are not part of any catalogue.}
\label{fig5_1}
\end{figure}

Recently, we have extended the sample of known X-ray cavities into the higher-redshift Universe \citep[$0.3<z<0.6$;][hereafter HL2012]{Hla2012421} using the MAssive Cluster Survey \citep[MACS;][]{Ebe2001553,Ebe2007661,Ebe2010407,Man2012420}. The MACS survey compiled the first large sample of very X-ray luminous clusters of galaxies at intermediate to high redshift, and consists of 124 spectroscopically confirmed clusters at $0.3 \leq{z}\leq{0.7}$ (see Fig. \ref{fig5_1}). Our work on the MACS clusters using $Chandra$ observations showed that X-ray cavities remain common in high-redshift cool-core clusters and that the energetics are still capable of preventing the surrounding gas from cooling. In particular, we found no evidence for evolution in any of the jetted outflow properties, implying that extreme mechanical AGN feedback has been in place for at least the past 5 Gyrs. We also noted that many of the clusters with X-ray cavities in the MACS sample had bright X-ray AGN. This is rarely seen in clusters of similar luminosities at low redshift \citep[][hereafter HL2011]{Hla2011}, where most of the X-ray luminous clusters at $z<0.3$ have no detectable X-ray nucleus, suggesting that we may be seeing some form of evolution in the radiative properties of these black holes. AGN feedback appears to be operating differently in high-redshift BCGs with X-ray cavities, at least in terms of the radiative properties of the AGN.  

Such an evolution is indicative of the predicted evolution between ``quasar-mode" and ``radio-mode" feedback. Local active SMBHs are mostly thought to be in radio-mode feedback, where the accretion rates are low and the black hole is capable of driving powerful jetted outflows. This explains the behaviour of nearby galaxy cluster cores ($z\aplt0.2-0.3$), which appear to have radiatively inefficient central AGN, yet are capable of driving large kpc scale jetted outflows. On the other hand, quasar-mode feedback is powered by a radiatively efficient black hole accreting at rates near the Eddington limit. This feedback mode has been invoked to explain the apparent deficit of extremely luminous galaxies in the galaxy luminosity function as well as the tight relations between black hole mass and host galaxy properties (e.g. stellar bulge mass and stellar velocity dispersion). Simulations have also demonstrated that quasar-mode feedback can regulate star formation, by heating and dispersing the cold star forming gas, thus terminating star formation \citep{DiM2005433,Cro2006365,Bow2006370,Spr2005435,Sij2006366,Mer2008388,Tey2011414,McC2011412,Mar2012420}. Although the quasar luminosity function peaks at high redshift \citep[$z\approx2-3$, e.g.][]{Wal2005434}, indicating that AGN in the past are more radiatively efficient, and that present day SMBHs are predominantly operating in a radiatively inefficient mode, it is still not clear how AGN transit from one mode to the other. In analogy with X-ray binaries, the current paradigm from the theoretical point of view is that the accretion disk transits from being geometrically thin in the radiatively efficient phase to being thick in the inefficient phase \citep{Chu2005363,Mer2008388}. Clusters of galaxies provide one of the most direct pieces of evidence for AGN feedback, with clear evidence that black holes can have a substantial impact on their surrounding medium (i.e. by the observations of X-ray cavities excavated by the AGN-driven outflow). By studying the radiative properties of the central AGN in clusters with known X-ray cavities (i.e. with known radio-mode feedback taking place) as a function of redshift, we can therefore provide insight into how AGN transit from quasar-mode to radio-mode feedback. 

The objective of this paper is therefore to further analyse the radiative properties of BCGs with X-ray cavities. We first describe our sample selection criteria in Section 2, and derive relevant cluster properties in Section 3. In Section 4, we measure the nuclear X-ray luminosities, and then the radio luminosities in Section 5. Section 6 presents the results and Section 7 discusses high-redshift clusters. Finally, in Section 8 we discuss the implications of our results and in Section 9, we present the conclusions. We adopt $H_\mathrm{0}=70\kmpspMpc$ with $\Omm=0.3$, $\OmL=0.7$ throughout this paper. All errors are $1\sigma$ unless otherwise noted.

\section{Sample selection}

Our aim is to study the radiative evolution of the AGN in BCGs with known X-ray cavities. We begin by selecting the intermediate to high redshift clusters ($0.3<z<0.6$) with known X-ray cavities, and base this on our previous work in HL2012, where we identified X-ray cavities in MACS using $Chandra$ images. $Chandra$ observations are required since it is the only X-ray telescope that has the resolving power to identify X-ray cavities beyond the local Universe ($z\apgt0.02$). 

76 MACS clusters have archival $Chandra$ observations. For each of these, we computed unsharp-masked and elliptical-subtracted X-ray images. Both techniques are used to enhance deviations in the original images. Using these processed images, we searched for X-ray cavities, and found 13 clusters with clear cavities (which could be clearly seen in the original images and processed images), as well as 7 with potential cavities (which could only be clearly seen in the processes images). Deeper observations are needed to confirm the potential cavities. We therefore only consider the first 13 as our high-redshift ($z>0.3$) sample of clusters with clear evidence for mechanical feedback from their central AGN. These are listed in Table \ref{tab5_1}. 

In order to study the radiative evolution of AGN feedback with cosmic time, we need to identify a comparison sample of low-redshift clusters with similar properties to the MACS clusters. We begin by searching the literature for clusters with known X-ray cavities at $z<0.3$. In total, we find 40 clusters with reported X-ray cavities. 
\begin{figure*}
\centering
\begin{minipage}[c]{0.99\linewidth}
\centering \includegraphics[width=\linewidth]{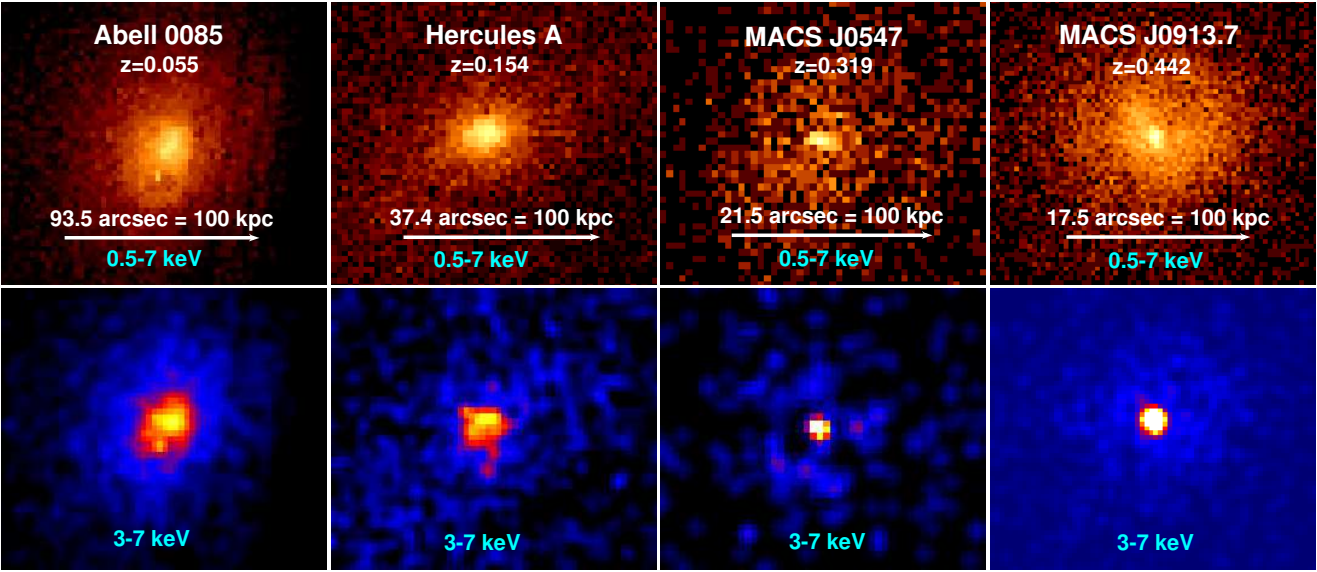}
\end{minipage}
\caption[$Chandra$ X-ray images for 4 clusters of galaxies in our sample]{Examples of 4 clusters of galaxies in our sample in ascending order of redshift from left to right. The top panels show the $0.5-7\keV$ images, whereas the bottom panels show the $3-7\keV$ images where the non-thermal nuclear emission from the central point source starts to dominate over the extended cluster emission with increasing redshift. }
\label{fig5_2}
\end{figure*}

The MACS survey is based on an X-ray flux-limited sample, and only comprises highly X-ray luminous clusters. The flux limit for MACS is $10^{-12}\ergpcmsqps$ in the $0.1-2.4\keV$ energy band, corresponding to $3\times10^{44}\ergps$ at $z=0.3$ (minimum redshift for MACS). MACS therefore only consists of clusters that have luminosities larger than $3\times10^{44}\ergps$. In order to compare this population of clusters to the low-redshift counterparts, we must only consider low-redshift clusters with luminosities above $3\times10^{44}\ergps$. This is because X-ray luminous clusters ($\Lx>10^{44-45}\ergps$) with strong cool cores, such as the MACS clusters with X-ray cavities, require extreme feedback from their central AGN to offset cooling of the surrounding medium ($L_{\rm mech}>10^{44-45}\ergps$), whereas less X-ray luminous clusters ($\approx10^{43}\ergps$), even with a strong cool core, only require some $10^{43}\ergps$ of feedback from their central AGN. This translates to smaller and less powerful X-ray cavities in the less luminous clusters. If we were to include these in our sample, we would not be probing the same outflows as in MACS. Out of the 40 clusters with clear X-ray cavities at $z<0.3$, we therefore only consider those that have similar luminosities to MACS ($\Lx>3\times10^{44}\ergps$ in the $0.1-2.4\keV$ energy band). We use the available total cluster fluxes in various surveys based on the $ROSAT$ All-Sky Survey \citep[$RASS$;][]{Ebe1996281,Ebe1998301,Ebe2000318,Ebe2002580,Boh2004425} to determine if a cluster is luminous enough to be included in our sample. If a source was present in more than one catalogue, we only considered the most recent catalogue. Our final sample is shown in Table \ref{tab5_1}, and examples in ascending order of redshift are shown in Fig. \ref{fig5_2}. The catalogue cluster fluxes are also shown in Table \ref{tab5_2}.

\begin{table}
\addtolength{\tabcolsep}{-2pt}
\caption[Sample of clusters]{Sample of luminous clusters with known X-ray cavities.}
\centering
\begin{tabular}{lcc}
\hline
\hline
(1) & (2) & (3) \\
Name & Alternate name& Redshift  \\
\hline
Perseus & Abell 0426 & 0.0183  \\
Abell 0085 & ... & 0.055  \\
Cygnus A & ... & 0.0561\\
Abell 1795 & ... & 0.063 \\
Abell 2029 & ... & 0.077  \\
Abell 2597 & ... & 0.085  \\
Abell 0478 & ... & 0.0881 \\
RXC J1558.3$-$1410& ... & 0.0970 \\
RXC J1524.2$-$3154& ... & 0.1028  \\
PKS 0745-19 & ... & 0.1028  \\
Abell 2204 & ... & 0.1522  \\
Hercules A & ... & 0.154 \\
Abell 0115 & ... & 0.1971 \\
ZwCl 2701 & ... & 0.2151  \\
MS0735.6+7421 & ZwCl 1370 & 0.2160 \\
4C+55.16 & ... & 0.2412 \\
Abell 1835 & ... & 0.2532  \\
ZwCl 3146 & ... &0.2906   \\
H1821+643 & ... & 0.299\\
MACS~J2140.2$-$2339 & MS 2137.3-2353 & 0.313  \\
MACS~J0242.5$-$2132 & ... & 0.314 \\
MACS~J0547.0$-$3904 & ... & 0.319  \\
MACS~J1931.8$-$2634 & ... & 0.352  \\
MACS~J0947.2+7623 & RBS 0797 & 0.354  \\
MACS~J1532.8+3021 & RX J1532.9+3021 & 0.3613   \\
MACS~J1720.2+3536 & Z8201 & 0.3913   \\
MACS~J0429.6$-$0253 & ... & 0.397 \\
MACS~J0159.8$-$0849 & ... & 0.404    \\
MACS~J2046.0$-$3430 & ... & 0.423   \\
MACS~J0913.7+4056 & IRAS 09104+4109 & 0.442   \\
MACS~J1411.3+5212 & 3C295 & 0.460   \\
MACS~J1423.8+2404 & ... & 0.5449  \\

\hline
\end{tabular}
\label{tab5_1}
\end{table}

Our sample also includes the powerful radio galaxy, 4C+55.16 ($z=0.2412$). 4C+55.16 is not part of any of the catalogues, but is embedded in a luminous cluster that has an X-ray luminosity within 200 kpc larger than many of the other clusters in our sample (see Table \ref{tab5_2}). It is therefore reasonable to assume that if it were part of the catalogues, its X-ray luminosity would be larger than $3\times10^{44}\ergps$, and therefore should be included in our sample.

Finally, we include H1821+643 ($z=0.299$), which is a quasar embedded in a luminous cool-core cluster \citep[][]{Rus2010402}. Since H1821+643 shows evidence for X-ray cavities that coincide with radio lobes, we included this object in the initial sample of 40 clusters with known X-ray cavities. H1821+643 is however not part of any of the catalogues, and this quasar is heavily affected by pile-up. However, \citet[][]{Rus2010402} have accounted for the pile-up by simulating the quasar PSF to disentangle the quasar emission from the intracluster medium. We therefore use the quantities quoted in \citet[][]{Rus2010402} throughout this paper instead of reanalyzing the source. \citet[][]{Rus2010402} furthermore estimate that the cluster luminosity in the $0.5-7\keV$ range is $\approx10^{45}\ergps$, of which most originates from the $0.1-2.4\keV$ energy band. It is therefore reasonable to assume that this cluster is luminous enough to be included in our sample.

\section{Cluster X-ray luminosities}

Our final sample comprises 32 clusters of galaxies in the $0.0<z<0.6$ redshift range. It includes some of the well known luminous clusters such as the Perseus cluster, MACS J1931.8-2634 \citep{Ehl2011411} and MS~0735.6+7421, the latter of which has the most powerful mechanical outburst known to date \citep[$10^{46}\ergps$;][]{McN2005433,Git2007660}. 

The cluster X-ray luminosities as derived from each of the catalogues are shown in column 4 of Table \ref{tab5_2}, and the catalogue references in column 8. As mentioned in the previous section, H1821+643 and 4C+55.16 are not part of the catalogues. However, \citet[][]{Rus2010402} estimate that the cluster luminosity for H1821+643 in the $0.5-7\keV$ range is roughly $10^{45}\ergps$. According to their spectra, most of this emission originates from the $0.1-2.4\keV$ energy band, and we therefore approximate that the $0.1-2.4\keV$ cluster luminosity for this source is $10^{45}\ergps$. For 4C+55.16, we compute a rough estimate of the cluster luminosity by fitting {\sc mekal} model to the $0.5-7\keV$ $Chandra$ emission within the central 200 kpc. The Galactic absorption is kept frozen at the \citet{Kal2005440} value and the background is taken as a region located on the same detector but far from any cluster emission.

\begin{table*}
\centering
\addtolength{\tabcolsep}{-1.pt}
\caption[Cluster properties]{Cluster properties - (1) Name; (2) $Chandra$ observation identification number; (3) Remaining exposure time of the $Chandra$ observations after data reduction; (4) X-ray luminosity in the $0.1-2.4\keV$ energy band as derived from catalogues, see Column 8; (5) Cooling radius, defined as the radius where the cooling time equals 7.7 Gyrs; (6) X-ray luminosity within the cooling radius in the $0.1-2.4\keV$ energy band; (7) Cavity enthalpy ($PV_{\rm cavities}$) taken from the literature; (8) Reference for the catalogue where the X-ray luminosity in Column 4 was taken from: i) Ebeling et al. in preparation.; ii) \citet{Boh2004425}; iii) \citet{Ebe2002580}; iv) \citet{Ebe2000318}; v) \citet{Ebe1998301}; vi) \citet{Ebe1996281}. $^a$Cluster luminosity derived as the $0.1-2.4\keV$ luminosity within 200 kpc. $^b$Values derived from \citet{Rus2010402}.}
\resizebox{17.6cm}{!} {
\centering
\hspace{-0.7cm}
\begin{tabular}{lccccccc}
\hline
\hline
(1) & (2) & (3) & (4) & (5) & (6) & (7) & (8) \\
Name & ObsID & Exp. & $\Lx{_{\rm , catalogue}}$ &  $r_{\rm cool}$ & $L_{\rm cool}$ & $PV_{\rm cavities}$ & Reference \\
& & (ks) & ($10^{44}\ergps$) & (kpc) & ($10^{44}\ergps$) & ($10^{58}\erg$)  & \\
\hline
Perseus cluster & 4952 & 149.9 & 7.03&  110 & 3.394$\pm$0.003 & 5.3 & (iii)\\
Abell 0085 & 904 & 37.1 & 5.2& 68 & 1.00$\pm$0.01 & 1.2 & (vi)\\
Cygnus A & 1707 & 9.2 & 4.0&  59 & 1.00$\pm$0.02& 84 & (iii)\\
Abell 1795 & 493+494 & 36.3 & 6.5&  89& 2.40$\pm$0.01 & 4.7 & (v)\\
Abell 2029 & 891 & 19.9 & 8.9&  81& 3.17$\pm$0.01 & 4.8 & (v)\\
Abell 2597  & 7329 & 59.1 & 4.7&  92& 2.39$\pm$0.01 & 3.6 & (vi)\\
Abell 0478 & 1669 & 40.3 & 7.7&  95& 5.34$\pm$0.02 & 1.5 & (v)\\
RXC J1558.3$-$1410 & 9402 & 36.4 & 3.4&  81& 1.46$\pm$0.01& ...& (ii)\\
RXC J1524.2$-$3154 & 9401 & 41.5 & 3.0&  78& 1.89$\pm$0.01& ...& (ii)\\
PKS 0745-19 & 12881 & 117.6 &12.3 &  101& 7.74$\pm$0.02 & 69 & (iii)\\
Abell 2204 & 7940 & 73.1 & 13.3 &  90& 7.97$\pm$0.03 & 4.0 & (vi)\\
Hercules A & 5796+6257& 96.2 & 3.6&  60& 0.63$\pm$0.01 & 31 & (v)\\
Abell 0115 & 3233 & 43.1 &10.0 & 80& 1.38$\pm$0.02 & 43 & (v) \\
ZwCl 2701 & 3195 & 25.9 & 7.6& 79& 1.91$\pm$0.03 & 350 & (v)\\
MS 0735.6+7421 & 10470 & 135.0 & 5.46&  79& 1.99$\pm$0.02 & 1600 & (iv)\\
4C+55.16 & 4940 & 73.6 & $\approx4.8^a$ & 91& 3.09$\pm$0.03 & 10 &...\\
Abell 1835 & 6880 & 113.8 & 28.7 & 100& 10.8$\pm$0.1 & 47&  (v)\\
ZwCl 3146 & 909+9371 & 77.8 & 20.6&  131& 12.8$\pm$0.1 & 380& (v)\\
H1821+643 & 9398+9845+9846+9848 & 85 & $\approx10^b$ & $\approx90^b$ & $\approx10^b$ & 14 & ...\\
MACS~J2140.2$-$2339 & 4974+5250+928&76.8 & 7.7  & 107& 8.13$\pm$0.05 &3.4 &(i)\\
MACS~J0242.5$-$2132 & 3266 & 8.6 & 10.1  & 110 & 9.51$\pm$0.2 & 9.0 & (i)\\
MACS~J0547.0$-$3904 & 3273& 19.2 &  5.9 &100 & 3.1$\pm$0.1 & 3.2 & (i)\\
MACS~J1931.8$-$2634 &  9382& 95.0& 12.6 & 112& 3.47$\pm$0.03 &83 & (i)\\
MACS~J0947.2+7623 &  7902 & 38.8 & 13.7 & 120& 15.3$\pm$0.1 &145 & (i)\\
MACS~J1532.8+3021 &  1649& 9.5 &  11.8  & 115& 13.6$\pm$0.3 &32 & (i)\\
MACS~J1720.2+3536 &  3280+6107+7718& 51.7& 9.7  & 100& 4.9$\pm$0.1 &18 &(i)\\
MACS~J0429.6$-$0253 & 3271 & 21.4 & 9.5 &  105& 6.5$\pm$0.2 & 2.2 & (i)\\
MACS~J0159.8$-$0849& 3265+6106+9376 & 61.4 & 11.5 & 86 & 6.0$\pm$0.1& 13 & (i) \\
MACS~J2046.0$-$3430 & 9377& 35.9& 6.9 &  81& 5.0$\pm$0.1 & 23 &(i)\\
MACS~J0913.7+4056 & 10445& 70.4 & 9.7 &  107& 8.8$\pm$0.1 & 150& (i) \\
MACS~J1411.3+5212 &  2254& 76.4& 6.7 &  60& 2.4$\pm$0.1 & 49 & (i)\\
MACS~J1423.8+2404 &  4195& 106.9 & 9.4 & 95 & 9.2$\pm$0.1 & 61 & (i)\\

\hline
\end{tabular}}
\label{tab5_2}
\end{table*}

\begin{figure}
\centering
\begin{minipage}[c]{0.99\linewidth}
\centering \includegraphics[width=\linewidth]{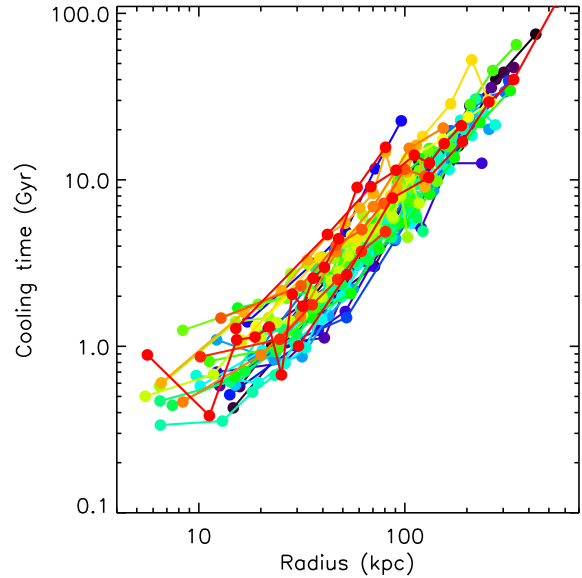}
\end{minipage}
\vspace{-0.2in}
\caption[Cooling time profiles]{Cooling time profiles for all of our clusters shown in Table \ref{tab5_1}. The cooling time profile for H1821+643 was taken from \citet{Rus2010402}. Deprojected thermal plasma parameters were used to derive the profiles. Each colour illustrates a different cluster. }
\label{fig5_3}
\end{figure}

We now focus on deriving the cooling luminosities in order to measure the power of the predicted cooling flows. We define the cooling luminosity as the $0.1-2.4\keV$ luminosity within the cooling radius, and we define the cooling radius using the same definition as \citet{Raf2006652}, i.e. the radius at which the cooling time ($t_{\rm cool}$) is equal to the $z=1$ look-back time. We adopt the same definition so that we can directly compare our results with theirs. For our cosmology, this corresponds to $t_{\rm cool}=7.7$ Gyrs. 
To determine the radius at which $t_{\rm cool}=7.7$ Gyrs, we have computed detailed cooling time profiles for all of our clusters. We use the same method as in HL2012, where $t_{\rm cool}$ is calculated with Eq. \ref{eq5_1}. Here $n_{\rm e}$ is the electron density, $kT$ is the gas temperature, $L_{\rm X}$ is the gas X-ray luminosity and $V$ is the gas volume contained within a shell.
\begin{equation}
t_{\rm cool} = \frac{5}{2}\frac{1.92~n_{\rm e}~kT~V}{L_{\rm X}}
\label{eq5_1}
\end{equation}

For each cluster, we selected the deepest observational $Chandra$ data set available. These data were then processed, cleaned and calibrated using the latest version of the {\sc ciao} software ({\sc ciaov4.4}, {\sc caldb4.4.9/10}), and starting from the level 1 event file. We applied both CTI (charge time interval) and time-dependent gain corrections, as well as removed flares using a $3\sigma$ threshold. We also exclude point sources during the fit. When a cluster was observed multiple times with the target centred on the same detector and in the same observing mode ({\sc faint } or {\sc vfaint}), we combined the different observations only if this would improve the image quality significantly. However, for MS 0735.6+7421 and the Perseus cluster, we only consider the deepest observation since they are sufficient to obtain detailed cooling time profiles. 

The thermal gas properties are then determined by selecting a set of annuli containing the same number of counts and centred on the X-ray peak. The total number of annuli depend on the data quality, but we chose the minimum number of counts so that we have at least 5 data points in the cooling time profile within 100 kpc, allowing us to estimate accurately the cooling radius when $t_{\rm cool}=7.7$ Gyrs. For the more nearby and well studied objects with deep $Chandra$ observations, we allowed several tens of thousand counts per annulus, but for the most distant and poorly observed ones, we could only allow some 2000 counts per annulus. The background region was chosen either within the same chip but far from any cluster emission or, for the more nearby objects, within a chip located on the same detector. Since we focus on the central 200 kpc of our clusters, where the emission remains strong, the choice of background does not significantly affect our results. We then proceeded in deprojecting the data using the Direct Spectral Deprojection method of \citet{San2007381}. For each deprojected spectrum, we fitted an absorbed (Galactic) {\sc mekal} model. In some annuli, it was difficult to constrain the abundance. In this case, we kept the abundance value frozen at the cluster average which we determined by selecting a region within the entire cluster ($r<200$ kpc). Point sources were excluded from the fitted region. The derived deprojected spectral quantities were then used as an estimate of the plasma parameters, allowing us to derive $t_{\rm cool}$ based on Eq. \ref{eq5_1}.  

\begin{table*}
\addtolength{\tabcolsep}{-1pt}
\caption[Core X-ray count rates, fluxes and luminosities]{Core X-ray count rates ($0.5-7\keV$), as well as observed fluxes and luminosities in the $2-10\keV$ energy band within a 6.4kpc$\times$6.4kpc square region. Internal nuclear absorption at the redshift of the source has not been considered here (see Table 4 for corrected luminosities). $^a$The X-ray nuclear luminosity for Perseus was taken from \citet{Mer2007381}. $^b$Values for H1821+643 were taken from \citet{Rus2010402}.}
\centering
\hspace{-0.7cm}
\begin{tabular}{lccc}
\hline
\hline
(1) & (2) & (3) & (4) \\
Name & Count rate & $\Fx{_, \rm core}$ & $\Lx{_, \rm core}$ \\
& (Cts s$^{-1}$) & ($10^{-14}\ergpcmsqps$) & ($10^{42}\ergps$) \\
\hline
Perseus cluster$^a$ & ... & ...  & 25$^{+54}_{-17}$  \\
Abell 0085 & 0.007$\pm$0.002  & 4.3$\pm$1.2  & 0.31$\pm$0.09  \\
Cygnus A & 0.307$\pm$0.007  & 131$\pm$3  &9.8$\pm$0.2 \\
Abell 1795 & 0.015$\pm$0.002  & 6.5$\pm$0.8  & 0.62$\pm$0.08  \\
Abell 2029 & 0.032$\pm$0.003  & 14.1$\pm$1.3  & 2.1$\pm$0.2 \\
Abell 2597  & 0.008$\pm$0.001  & 3.4$\pm$0.5  & 0.61$\pm$0.08  \\
Abell 0478 & 0.006$\pm$0.001  & 3.9$\pm$0.9  & 0.8$\pm$0.2 \\
RXC J1558.3$-$1410 & 0.0055$\pm$0.0008  & 2.3$\pm$0.3  & 0.55$\pm$0.08  \\
RXC J1524.2$-$3154 & 0.004$\pm$0.001  & 1.7$\pm$0.5  & 0.4$\pm$0.1  \\
PKS 0745-19 & 0.0005$\pm$0.0003  & 2.4$\pm$1.6  & 0.6$\pm$0.4  \\
Abell 2204 & 0.0044$\pm$0.0007  & 3.1$\pm$0.5  & 1.9$\pm$0.3  \\
Hercules A & 0.0029$\pm$0.0003  & 1.4$\pm$0.2  & 0.91$\pm$0.09  \\
Abell 0115 & 0.0004$\pm$0.0002  & 0.3$\pm$0.2  & 0.3$\pm$0.2  \\
ZwCl 2701 & 0.0005$\pm$0.0003  & 0.2$\pm$0.1  & 0.3$\pm$0.2 \\
MS 0735.6+7421 &0.0003$\pm$0.0001  & 0.22$\pm$0.09  & 0.3$\pm$0.1  \\
4C+55.16 & 0.0105 $\pm$0.0005  & 4.9$\pm$0.2  & 8.6$\pm$0.4  \\
Abell 1835 & 0.0006$\pm$0.0003  & 0.4$\pm$0.2  & 0.8$\pm$0.4  \\
ZwCl 3146 &0.0021 $\pm$0.0003  & 1.5$\pm$0.2  & 3.9$\pm$0.6  \\
H1821+643$^b$ & ... & ...  & 4200$^{+100}_{-100}$  \\
MACS~J2140.2$-$2339 & 0.0029$\pm$0.0003  & 1.3$\pm$0.1  & 4.3$\pm$0.5  \\
MACS~J0242.5$-$2132 & 0.0026$\pm$0.0009  & 1.8$\pm$0.6  & 5.8$\pm$2.0  \\
MACS~J0547.0$-$3904 & 0.0037$\pm$0.0005  & 2.5$\pm$0.3  & 8.4$\pm$1.1  \\
MACS~J1931.8$-$2634 & 0.0050$\pm$0.0002  & 3.7$\pm$0.2  & 15.3$\pm$0.7  \\
MACS~J0947.2+7623 & 0.0127$\pm$0.0006  & 5.7$\pm$0.2  & 24.0$\pm$1.1 \\
MACS~J1532.8+3021 & 0.0003$\pm$0.0002  & 0.15$\pm$0.09  & 0.7$\pm$0.4  \\
MACS~J1720.2+3536 & 0.0004$\pm$0.0002  & 0.3$\pm$0.1  & 1.6$\pm$0.6  \\
MACS~J0429.6$-$0253 & 0.0011$\pm$0.0004  & 0.8$\pm$0.3  & 4.4$\pm$1.6  \\
MACS~J0159.8$-$0849 & 0.0016$\pm$0.0003 & 1.1$\pm$0.2  & 6.2$\pm$1.1  \\
MACS~J2046.0$-$3430 & 0.0007$\pm$0.0003  & 0.5$\pm$0.2  & 3.3$\pm$1.2  \\
MACS~J0913.7+4056 & 0.0070$\pm$0.0004  & 4.7$\pm$0.3  & 33.6$\pm$2.0  \\
MACS~J1411.3+5212 & 0.0038$\pm$0.0003  & 2.5$\pm$0.2  & 20.0$\pm$1.5  \\
MACS~J1423.8+2404 & 0.0008$\pm$0.0002  & 0.33$\pm$0.07  & 3.9$\pm$0.8  \\
\hline
\end{tabular}
\label{tab5_3}
\end{table*}

Once we obtained the cooling time profiles (Fig. \ref{fig5_3}), we determined the cooling radius, defined as the radius when $t_{\rm cool}=7.7$ Gyrs. The cooling luminosity ($L_{\rm cool}$) then corresponds to the $0.1-2.4\keV$ luminosity within this radius. We show our results in Table \ref{tab5_2}, Column 6. For H1821+643, we use the cooling time profile derived by \citet{Rus2010402} to estimate the cooling radius when $t_{\rm cool}=7.7$ Gyrs ($\approx90$ kpc), and then estimate the cooling luminosity based on the average heating required within the cooling radius as derived by  \citet[][; $10^{43}\ergps$kpc$^{-1}$]{Rus2010402}.

\section{Nuclear X-ray luminosities}

$Chandra$ X-ray observations are used to isolate the nuclear X-ray emission for our sample of BCGs. Some are dominated by non thermal emission, while others have no detectable X-ray nucleus (e.g. A1835, A478, Z2701). We therefore derive the nuclear luminosities using 2 methods. The first is used to estimate the fluxes in a systematic way for all objects, regardless of the nuclear spectrum and is based only on the core count rate. The second focusses on the objects with clear power-law components in their spectrum, and we determine their nuclear luminosities by fitting a non-thermal model to the spectra. The two methods are described in the following sections. 
\begin{figure}
\centering
\begin{minipage}[c]{0.99\linewidth}
\centering \includegraphics[width=\linewidth]{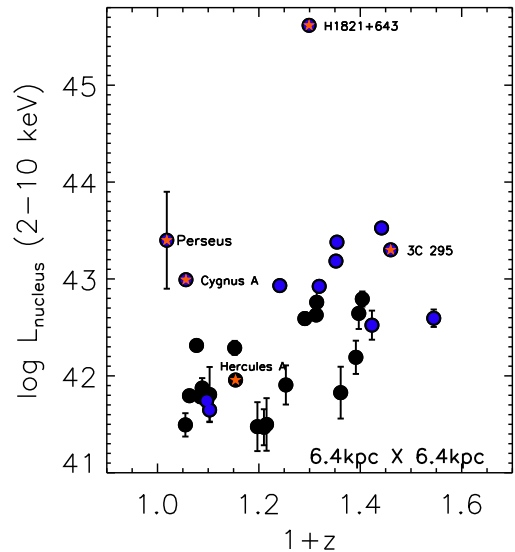}
\end{minipage}
\caption[Core X-ray luminosities as a function of redshift, based on count rates]{Logarithm of the $2-10\keV$ X-ray core luminosity in $\ergps$ as a function of redshift, derived from the observed count rates (i.e. not corrected for any internal absorption, if present). The Perseus cluster, Hercules A, H1821+643, Cygnus A and 3C 295 are highlighted with the red stars. Note that the values for Perseus and H1821+643 have been taken from the literature and have therefore already been corrected for internal absorption. The blue points illustrate the sources with evidence of nuclear non-thermal emission associated with the BCG. }
\label{fig5_4}
\end{figure}

\subsection{Nuclear luminosities based on a core count rate}

For the first method, we use the web interface of {\sc pimms}\footnote[1]{http://heasarc.gsfc.nasa.gov/Tools/w3pimms.html} \citep{Mukai1993}, which allows us to convert a count rate into a flux. We first consider the highest redshift object in our sample, MACS~J1423.8+2404 located at $z=0.5449$ ($1''=6.4$kpc). We select a 2 by 2 pixels square region in the $0.5-7\keV$ $Chandra$ image, which corresponds to a $1''\times1''$ (or 6.4kpc$\times$6.4kpc) square region. Then, we select a surrounding region as a background, and more precisely, we consider a square annulus centred on the same position, with an outer 6 by 6 pixels square and an inner 4 by 4 pixels square. The total count number of the core is then calculated by subtracting the background emission scaled to the same number of pixels as the central 2 by 2 pixels square. {\sc pimms} is then used to convert the count rate into a flux, and then a luminosity in the $2-10\keV$ energy band. A power-law with photon index of 1.9 is used as a model, but our results are not sensitive to our value of the index, at least within $\pm0.2$. We then proceed with the other objects, and select a central 6.4kpc$\times$6.4kpc square for each cluster. The background is selected in a similar way, as a surrounding square annulus. Since X-ray counts are governed by Poisson noise, we can estimate that the error associated with each square region is given as the square root of the number of counts. Using the error propagation equation, we then estimate the 1$\sigma$ noise level for the background-subtracted count rate which we show in Table \ref{tab5_3}.

We consider the same region for all objects in terms of kpc$^2$, and not in terms of arcsec$^2$. This is because a $1''\times1''$ square region at high redshift is sampling more of the cluster emission than a $1''\times1''$ square region at low redshift. For the nearest objects, the central 6.4kpc$\times$6.4kpc square region therefore extends well over a dozen pixels. We choose this particular dimension because we are limited by the Point Spread Function (PSF) of the telescope for the high-redshift objects. For a $1''\times1''$ square region, some 70 per cent of the energy of a point source falls in. If we were to choose an even smaller region, we would be missing most of the nuclear fluxes for our high-redshift objects. Although this method does have its limitations, it should only affect the high-redshift objects by slightly underestimating their fluxes, and the low-redshift objects by slightly overestimated their fluxes since it will be including the counts from the nucleus and surrounding thermal emission. This means that any evolution we find, is probably a lower limit to the true evolution, which should be even more rapid. Our results are illustrated in Fig. \ref{fig5_4}.

Note that, we could not derive a core flux for Perseus or H1821+643 based on the $Chandra$ observations which are affected by pileup. Pileup occurs when two or more photons are detected as one event \citep[see for more details][]{Dav2001562,Rus2009402}, and the amount of pileup can be estimated by comparing the fraction of good (grades 0,2,3,4,6) to bad grades (grades 1,5,7) for each point source. Typically, pileup becomes problematic when the fraction of bad grades exceeds 10 per cent of the good grades. For Perseus, the fraction of bad grades exceeds 15 per cent even for the observations taken with a reduced frame rate (ObsID 3404), and for H1821+643, the observations are heavily affected by pileup such that a readout streak is seen. Instead, we use the nuclear flux quoted in \citet{Rus2010402} for H1821+643 and in \citet{Mer2007381} for Perseus, the latter of which is an average estimate based on the fluxes available in the literature. The nuclear flux of Perseus also appears to have varied significantly over the last 10 years, according to the different values published in the literature \citep[see][]{All2001322,Eva2006642,Chu2003590,Don2004617}. To account for the order of magnitude variability, we assign an uncertainty of 0.5 in logarithm space ($\log\Lx=43.4\pm0.5$). 

Applying a similar method, this time for a 12.7kpc$\times$12.7kpc square region which comfortably contains $Chandra$'s PSF, we find that the scatter increases slighty, but that our conclusions discussed in Section 8 remain the same. We also investigate another method for deriving the nuclear fluxes. It consists of fitting a regression to the surface brightness profile within the central tens of kpc, for example a power-law, and then attributing the extra number of counts in the nuclear region as those associated with the central point source. The observed additional counts can then be converted to a nuclear flux \citep{All2006372}. However, as mentioned in HL2011, we investigated this possibility for many of our objects at low redshifts, but this method predicted a higher number of counts than that found within the nucleus. This is because many of our clusters have surface
brightness profiles that flatten strongly within the inner regions,
and this extrapolation approach cannot be used to estimate
the flux of the nucleus.

\begin{table*}
\addtolength{\tabcolsep}{-1pt}
\caption[]{Spectral modelling for the sources with evidence of non-thermal emission. (1) Cluster name; (2) Model name; (3) Absorbing column density at the redshift of the source; (4) Power-law index; (5) Gaussian central energy; (6) Gaussian dispersion; (7) Parameter for each model; (8) Unabsorbed nuclear X-ray luminosity in the $2-10\keV$ energy band. Galactic absorption was included for all sources, and was frozen at the values of \citet{Kal2005440}. $^a$Covering fraction of the absorber. $^b$Temperature and abundance values for the {\sc mekal} model. $^c$Power-law index of the first, unabsorbed power-law (at soft X-rays). $^d$X-ray nuclear luminosity for the central galaxy in the Perseus cluster, taken from \citet{Mer2007381}. $^e$Values for H1821+643 were taken from \citet{Rus2010402}. }
\centering
\resizebox{17.5cm}{!} {
\hspace{-0.3in}
\begin{tabular}{llcccccc}
\hline
Name & Model & ${}N_{\rm H}$ & $\Gamma$ & $E_{\rm gaussian}$ & $\sigma_{\rm gaussian}$ & Param. & $L_{\rm X,nuc}(2-10)\keV$ \\
& & ($10^{22}$cm$^{-2}$) & & (keV) & (eV) &  & ($10^{42}\ergps$) \\
\hline
(1) & (2) & (3) & (4) & (5) & (6) & (7) & (8) \\
\hline
Perseus cluster & ... & ... &...  &...  & ... & ... & $^d$25$^{+54}_{-17}$\\ 
& & & & & & & \\
Cygnus A & I:~~ phabs [ pow + zphabs(pow+ga)]&17.6$^{+3.4}_{-3.0}$ & $1.5^{+0.5}_{-0.5}$& 6.1& 50& [1.6$^{+0.5}_{-0.5}$]$^c$ & 173$^{+29}_{-19}$\\
&II:~ phabs [ mekal + zphabs(pow+ga)]& $16.5^{+2.9}_{-2.7}$&1.3$^{+0.5}_{-0.5}$ & 6.1& 50& [3.3$\keV$; 3.2Z$_{\rm \odot}$]$^b$& 167$^{+22}_{-16}$\\
& & & & & & & \\
RXC J1558.3$-$1410 & I:~~phabs [ pow + zphabs(pow) ] & $15^{+5}_{-3}$ & 1.9 & ... & ... & [$2.3^{+1.1}_{-1.0}$]$^c$ & $5.0^{+1.3}_{-1.0}$  \\ 
& II: phabs [ mekal + zphabs(pow) ] & $19^{+4}_{-3}$ & $1.9$ & ... & ... & [2.2$\keV$; 0.60Z$_{\rm \odot}$]$^b$ & $6.7^{+1.4}_{-1.1}$  \\ 
& & & & & & & \\
RXC J1524.3$-$3154 & I:~~phabs [ pow ] & ... & $1.5^{+0.2}_{-0.2}$ & ... & ... & ... & $0.8^{+0.2}_{-0.2}$  \\
& II: phabs [ zphabs(pow) ] & $1.5^{+0.9}_{-0.7}$ & $3.4^{+1.0}_{-0.8}$ & ... & ... & ... & $0.5^{+0.2}_{-0.2}$  \\
& & & & & & & \\
4C+55.16& I:~~phabs [ pow ] & ... & $1.54^{+0.07}_{-0.07}$ & ... & ... & ... & $12.0^{+1.1}_{-1.1}$\\
& & & & & & & \\
H1821+643& ... &...  & ... & ... & ... & ... & $^e$4200$^{+100}_{-100}$\\
& & & & & & & \\

MACS J0547.0$-$3904 & I:~~phabs [ pow ] & ... & $2.3^{+0.6}_{-0.5}$ & ... & ... & ... & $5.9^{+6.4}_{-3.0}$ \\
& II: phabs [ mekal + zphabs(pow) ] & $0.6^{+1.4}_{-0.5}$ & $2.2^{+0.7}_{-0.7}$ & ... & ... & [1.4$\keV$; 0.43Z$_{\rm \odot}$]$^b$ & $11.0^{+5.5}_{-3.7}$ \\
& & & & & & & \\
MACS J1931.8$-$2634& I:~~phabs [ zphabs(pow) ] & $1.8^{+0.4}_{-0.4}$ & $1.7^{+0.2}_{-0.2}$ & ... & ... & ... & $53.2^{+4.0}_{-3.7}$  \\
& & & & & & & \\
MACS J0947.2+7623& I:~~phabs [ zpcfabs(pow) ] & $5.3^{+0.8}_{-0.9}$ & 1.6$^{+0.1}_{-0.2}$ & ... &... & [$0.95^{+0.02}_{-0.02}$]$^a$ & $180^{+11}_{-11}$ \\
& II: phabs [ zpcfabs( pow + ga) ] &  $4.5^{+1.0}_{-0.9}$ & $1.4^{+0.4}_{-0.4}$ &  $3.0^{+0.1}_{-0.1}$ & $100$ & [$0.93^{+0.04}_{-0.02}$]$^a$ & $195^{+23}_{-26}$ \\
& & & & & & & \\
MACS J2046.0$-$3430 & I:~~phabs [ mekal + pow ] & ... & 1.9 & ... & ...& [1.8$\keV$; 0.39Z$_{\rm \odot}$]$^b$ & $2.9^{+2.2}_{-1.8}$  \\
& II: phabs [ mekal + zphabs(pow) ] & $15^{+31}_{-10}$ & 1.9 & ... & ... & [1.8$\keV$; 0.39Z$_{\rm \odot}$]$^b$ & $5.9^{+7.4}_{-3.3}$\\
& & & & & & & \\
MACS J0913.7+4056& I:~~phabs [ zpcfabs(pow + ga) ]& $48^{+18}_{-13}$ & $1.4^{+0.3}_{-0.4}$ & $4.4$ & $60$ & [$0.92^{+0.04}_{-0.07}$]$^a$ & $517^{+181}_{-157}$  \\
& II: phabs [ pow + zphabs(pow + ga) ] & $39^{+26}_{-19}$ & $1.1^{+1.0}_{-0.9}$ & $4.4$ & $60$ & [1.6]$^c$ & $450^{+290}_{-120}$  \\
& & & & & & & \\
MACS J1411.3+5212& I:~~~phabs [ zpcfabs(pow) ]& $44^{+3}_{-3}$ & $2.1^{+0.1}_{-0.1}$ & ... & ... & [$0.98^{+0.01}_{-0.01}$]$^a$  & $407^{+55}_{-60}$ \\
	& II:~~phabs [ pow + zphabs(pow) ] & $52^{+11}_{-10}$ & $2.5^{+0.5}_{-0.5}$ & ...& ...& [$1.6^{+0.5}_{-0.5}$]$^c$ & $490^{+168}_{-106}$  \\
	& III: phabs [ mekal + zphabs(pow) ] & $48^{+9}_{-9}$ & $2.4^{+0.5}_{-0.5}$ & ... & ... & [$3.7^{+49.7}_{-1.5}$; 0.62Z$_{\rm \odot}$]$^b$ & $456^{+139}_{-93}$ \\
	& & & & & & & \\
MACS J1423.8+2404& I:~~phabs [ pow ] & ... & $2.1^{+0.2}_{-0.2}$ & ...  & ...  & ...  & $8.2^{+2.3}_{-2.1}$  \\
& II: phabs [ mekal + pow ] & & $2.0^{+0.1}_{-0.2}$ & ...& ... & [3.9$\keV$; 0.65Z$_{\rm \odot}$]$^b$ & $14.3^{+6.9}_{-6.7}$  \\
& & & & & & & \\
\hline
\end{tabular}}
\label{tab5_5}
\end{table*}
\begin{figure*}
\centering
\hspace{-1cm}
\begin{minipage}[c]{0.55\linewidth}
\centering \includegraphics[width=\linewidth]{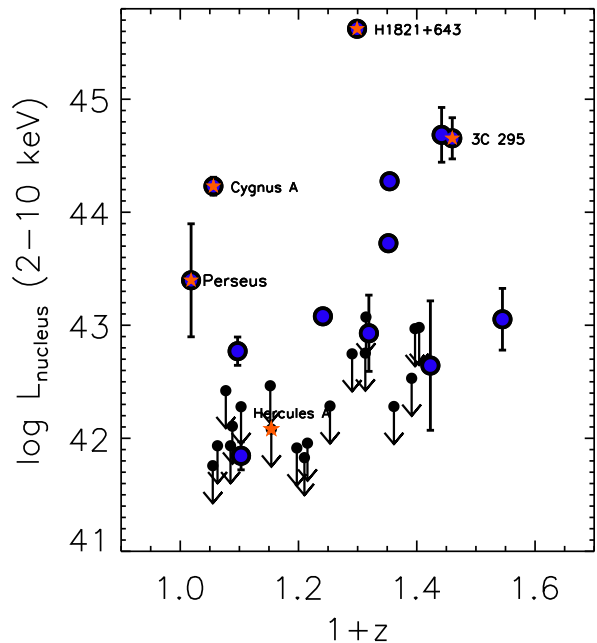}
\end{minipage}
\caption[X-ray nuclear luminosities as a function of redshift, based on model fitting]{Logarithm of the $2-10\keV$ nuclear luminosity as a function of redshift for all BCGs in our sample (in units of $\ergps$). The 13 objects that have nuclear luminosities derived from detailed model fitting are shown with the blue circles, whereas the 19 objects that have no detectable X-ray nucleus are shown with the downward pointing arrow ($3\sigma$ upper limit). The Perseus cluster, Hercules A, H1821+643, Cygnus A and 3C 295 are highlighted with the stars.  }
\label{fig5_4c}
\end{figure*}

\subsection{Nuclear luminosities based on spectral model fitting}

In the previous section, we obtained the core X-ray fluxes based on the observed count rates. This allowed us to obtain the fluxes in a systematic way for all clusters, which is especially useful for those that have no detectable X-ray nucleus. In this case, the background-subtracted count rate derived in the previous section could still include some thermal contribution and the luminosities obtained should be regarded as a conservative estimate of the upper limits of the non-thermal contribution. However, 13 of our objects show evidence of a central point source associated with the BCG (see Table \ref{tab5_5}), and in some cases, there is evidence of significant absorption in the nuclear spectra. The true unabsorbed nuclear X-ray luminosities for these 13 sources might therefore be significantly higher than those calculated in the previous section. Note that these 13 sources include Perseus and H1821+643. As mentioned in the previous section, we use the nuclear flux quoted in \citet{Mer2007381} for Perseus, and for H1821+643 we use the one from \citet{Rus2010402} throughout this paper.

\begin{table*}
\caption[Cluster radio properties]{Radio properties - (1) Name; (2) 74 MHz radio flux density from VLSS; (3) 326 MHz radio flux density from WENSS or \citet{Bir2008686}; (4) 1400 MHz radio flux density; (5) 5 GHz radio flux density; (6) 28.5 GHz radio flux density from Berkeley-Illinois-Maryland Association \citep[BIMA;][]{Cob2008134}; (7) Other radio flux density; (8) Integrated radio luminosity from 10 MHz to 10 GHz; (9) References: (i) 1.4 GHz FIRST; (ii) 1.4 GHz NVSS; (iii) Parkes MIT-NRAO 4.85 GHz survey \citep{Gri1993105}; (iv) 5 GHz private communication (M. T. Hogan); (v) 5 GHz from \citet{Hin1993415}; (vi) 5 GHz from \citet{Cav2011732}; (vii) 5 GHz from \citet{Gre199175}; (viii) EMSS 5 GHz \citep{Gio199072,Sto199176}; (ix) 843 MHz SUMSS \citep[][]{Boc1999117,Mau2003342}; (x) 15 GHz Arcminute Microkelvin Imager (AMI), private communication (K. Grainge); (xi) 1.4 GHz from \citet{Ehl2011411}; (xii) 150 MHz TIFR GMRT Sky Survey; (xiii) Radio fluxes at 327 MHz, 1.4 GHz, 4.5 GHz and 8.5 GHz, as well as the total radio luminosities derived in \citet{Bir2008686}. $^{a}$For the central galaxies of MACS~J0242.5$-$2132, MACS~J1411.3+5212 (3C 295), PKS 0745-19, 4C+55.16, A2029 and Hercules A, we use the extensive follow-up radio observations available in the NASA/IPAC Extragalactic Database (NED) to determine $L_{\rm radio}$. $^{b}$Abell 2029 and Hercules A have extended morphologies in the different surveys due to resolved radio lobes. The fluxes shown are the total integrated ones for all emission associated with the central AGN, including the contribution of the radio lobes.  }
\centering
\resizebox{17.5cm}{!} {
\hspace{-0.3in}
\begin{tabular}{lccccccccc}
\hline
\hline
(1) & (2) & (3) & (4) & (5) & (6) & (7) & (8) & (9) \\
Cluster name& $S_{74{\rm MHz}}$ & $S_{326/327{\rm MHz}}$ & $S_{1.4{\rm GHz}}$ & $S_{4.5/4.85{\rm GHz}}$ & $S_{28.5{\rm GHz}}$ & $S_{\rm other}$ & $L_{\rm radio}$ & Ref. \\
 & (mJy) & (mJy)  & (mJy)  & (mJy)  & (mJy)  & (mJy) & ($10^{42}\ergps$) & \\
\hline
Perseus cluster & ... & 24500$\pm$1000 & 23200$\pm$900 &... &... & 23900$\pm$1000[@8.5GHz]& 0.365& (xiii)\\
Cygnus A &  ... & 4375000$\pm$194000 & 1450000$\pm$60000 & 475000$\pm$20000&... & 180000$\pm$10000[@8.5GHz]& 690& (xiii\\
Abell 0085 & 1130$\pm$420& ...& 40.2$\pm$2.0& 46$\pm$11& ...& ... & 0.112&  (i,iii)\\
Abell 1795 & ... & 3360$\pm$140 & 880$\pm$4 &... &... & 99$\pm$4[@8.5GHz]& 0.70& (xiii)\\
Abell 2029$^{a,b}$ & 16770$\pm$190& ...& 526$\pm$14& 65$\pm$11& ...& ... & 1.66& (i,iii)\\
Abell 2597  & ... & 8300$\pm$300 & 1860$\pm$70 &370$\pm$20 &... & 118$\pm$5[@8.5GHz]& 3.1& (xiii)\\
Abell 0478 & ... & 110$\pm$10 & 27$\pm$1 &... &... & 5.4$\pm$0.2[@8.5GHz]& 0.029& (xiii)\\
RXC J1558.3$-$1410 & $1950\pm270$& ...& 461$\pm$23& 652$\pm$36&...& 1787$\pm$91[@150MHz] & 1.73&  (ii,iii,xii)\\
RXC J1524.2$-$3154 & ...&... & 49.8$\pm$2.5& ...& ...& 118$\pm$40[@150MHz] & 0.16& (ii,ix,xii)\\
& & & & & & 47.4$\pm$2.1[@843MHz] & ...&...  \\
PKS 0745-19$^{a}$ & 39060$\pm$200 & ...& 2372$\pm$119& 480$\pm$27& ...& 26770$\pm$1350[@150MHz] & 6.07& (ii,iii,xii)\\
Abell 2204 & 1640$\pm$290 & ...&  57.9$\pm$2.9& 25.9$\pm$1.3& 8.79$\pm$0.13& ... & 1.23&  (i,iv)\\
Hercules A$^{a,b}$ & 1038200$\pm$300 & ...& 47120$\pm$2356 & 11376$\pm$592&... & ... & 313.7 & (ii) \\
Abell 0115 & 38810$\pm$304& ...& 1362$\pm$68& ...&... & ... & 48.4&  (ii)\\
ZwCl 2701 & ... & 210$\pm$10 & ... &4.3$\pm$0.2 &... & 2.9$\pm$0.1[@8.5GHz]& 0.40& (xiii)\\
MS 0735.6+7421 & ... & 800$\pm$30 & 21$\pm$1 &... &... & 1.3$\pm$0.1[@8.5GHz]&9.89 & (xiii)\\
4C+55.16$^{a}$ & 11510$\pm$340 & 9184$\pm$459& 8240$\pm$412& ...&...& ... & 102.8&(i)\\
Abell 1835 & ... & 95$\pm$7 & 31$\pm$1 &9.9$\pm$0.4 &... & 7.4$\pm$0.3[@8.5GHz]& 0.37& (xiii)\\
ZwCl Z3146 & ... & 28$\pm$3 & ... &1.39$\pm$0.07 &... & 0.78$\pm$0.05[@8.5GHz]& 0.15& (xiii)\\
H1821+643 & 1140$\pm$840 & 602$\pm$24 & 91.9$\pm$4.7 & ... & ... & ... & 3.2 & (ii) \\
MACS~J2140.2$-$2339 & $<116$ & ...& $3.8\pm0.5$ & $1.0\pm0.1$ & ... & ... & $0.42$ &   (ii,viii)\\
MACS~J0242.5$-$2132$^{a}$ & $890\pm145$     & ... & $1255\pm73$ & $795\pm43$ & ... & ... & 27.1 &   (ii,iii)\\
MACS~J0547.0$-$3904 & ... & ...                 & $31.4\pm1.9$ & $15.4\pm0.8$ & ... & $19.6\pm1.3$[@843MHz] & 0.59 &   (ii,iv,ix)\\
MACS~J1931.8$-$2634 & ... & ...& $70\pm4$ & $6.0\pm1.3$ & ... & $2799\pm161$[@150MHz] & 80.4 &   (iv,xi,xii)\\
MACS~J0947.2+7623 & $<329$ & $91.0\pm6.1$     &$21.7\pm1.3$ & $4.0\pm0.2$ & ... & ... & $1.02$ &   (ii,vi)\\
MACS~J1532.8+3021 & $<222$ & $71.0\pm8.2$ &$17.1\pm0.9$ & $8.8\pm0.5$ & $3.25\pm0.18$ & ... & $0.94$ &  (i,iv)\\
MACS~J1720.2+3536 & $<266$ & $103.0\pm7.4$ &$16.8\pm1.0$ & ... & ... & ... & $1.23$ &   (i)\\
MACS~J0429.6$-$0253 & $<214$ & ...              & $138.8\pm8.1$ & ... & ... & ... &  $7.2$ &  (ii)\\
MACS~J0159.8$-$0849 & $<112$ & ...              & $31.4\pm1.6$ & $58\pm11$  & ... & ... & $3.56$ &   (i,iii) \\
MACS~J2046.0$-$3430 & ... & ...& $8.1\pm0.6$ & ... & ... & $13\pm1.3$[@843MHz] & 2.49 &   (ii,ix)\\
MACS~J0913.7+4056 & $<272$ & $54.0\pm7.1$     &$8.3\pm0.5$ & $1.6\pm0.1$ & $0.69\pm0.12$ & $0.80\pm0.04$[@15GHz] & $1.33$  & (i,v,x)\\
MACS~J1411.3+5212$^{a}$ & $120270\pm6022$ & $61647\pm3082$ &$22171\pm1109$ & $7401\pm808$ & ... & ... & 1025.3 &  (i,vii)\\
MACS~J1423.8+2404 & $<232$ & ... &$5.2\pm0.4$ & ... & $1.49\pm0.12$ & ... & $3.58$ &   (i)\\


\hline
\end{tabular}}
\label{tab5_4}
\end{table*}

The spectral models adopted for each of the 13 sources, as well as the resulting, unabsorbed $2-10\keV$ nuclear luminosities are shown in Table \ref{tab5_5} (see Appendix A for the details of the modelling). For each source in Table \ref{tab5_5}, we calculate the average luminosity across the different models fitted and the associated uncertainty is calculated as the quadratic sum of the errors. For the latter, we assume that the errors are symmetric and only consider the average between the upper and lower bound uncertainty. For the remaining 19 objects that have no detectable X-ray nucleus, we take the $3\sigma$ upper bound of the $2-10\keV$ luminosities shown in Table \ref{tab5_3} (Column 4) as the upper limit to the non-thermal emission. Combining these 19 non detections with the 13 detections of Table \ref{tab5_5}, we illustrate the new evolution in Fig. \ref{fig5_4c}. 

To illustrate the steepness of the evolution, we perform a simple linear regression in the log-log space such that $L_{\rm nucleus}\propto(1+z)^\alpha$. We stress that the scatter in our plots is large and various selection effects may be present (see Section 6). We therefore cannot derive a precise estimate of how these BCGs evolve with time. We only use the regression to demonstrate that the AGN in BCGs with X-ray cavities are on average brighter at higher redshift. We use the method of \citet{Kel2007}\footnote[4]{The {\sc linmix$\_$err.pro} script is available at the IDL astronomy library, http://idlastro.gsfc.nasa.gov/.}, based on a Bayesian approach that uses a Monte Carlo technique to simulate the linear regression parameters from their probability distribution given the observed data. Note that upper limits can be incorporated into the fit, but these influence strongly the fit and drag the normalization downwards. Applying this method, we find that the slope of the regression is positive to more than a 95 per cent confidence level (i.e. $\apgt2\sigma$).

\begin{figure*}
\centering
\begin{minipage}[c]{0.49\linewidth}
\centering \includegraphics[width=\linewidth]{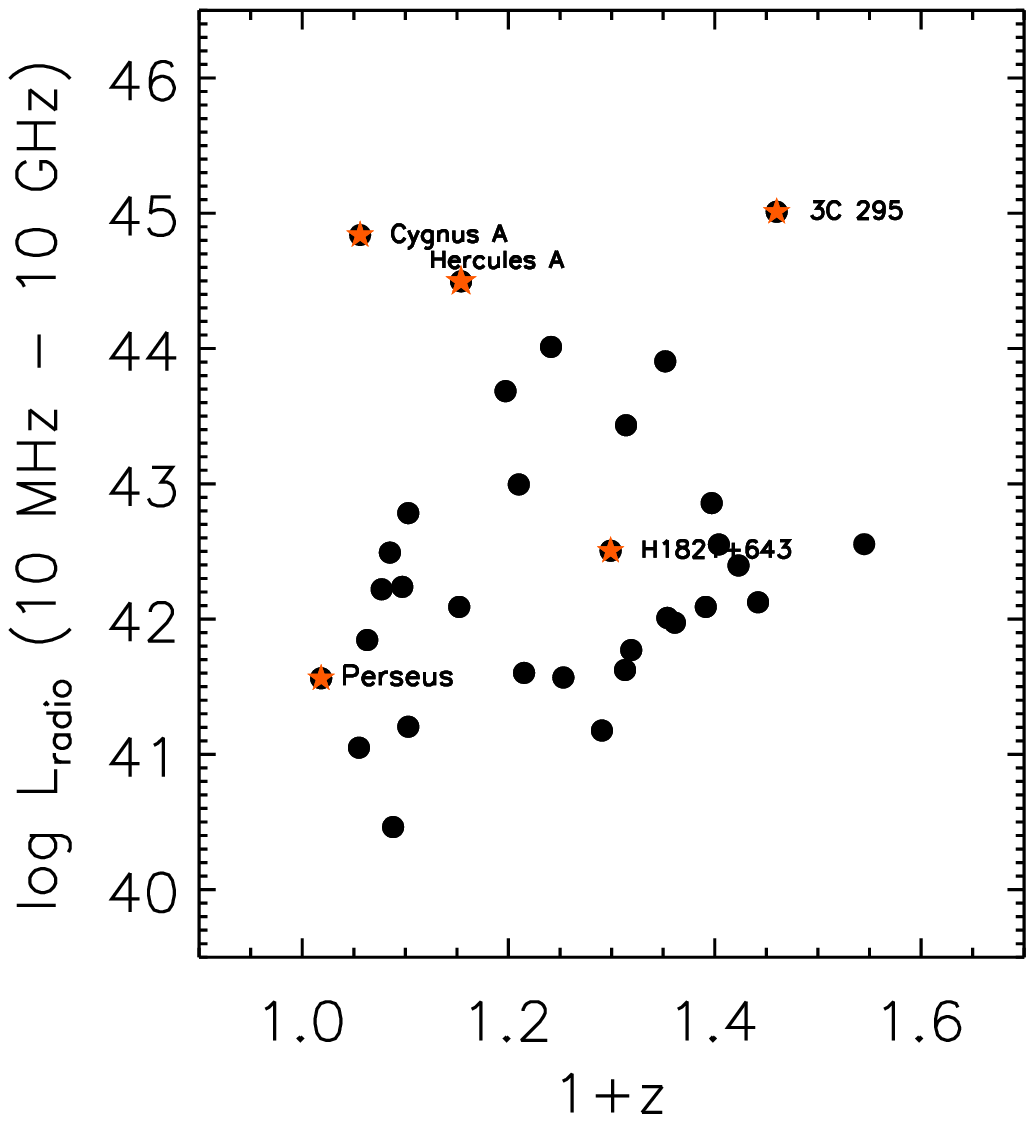}
\end{minipage}
\begin{minipage}[c]{0.49\linewidth}
\centering \includegraphics[width=\linewidth]{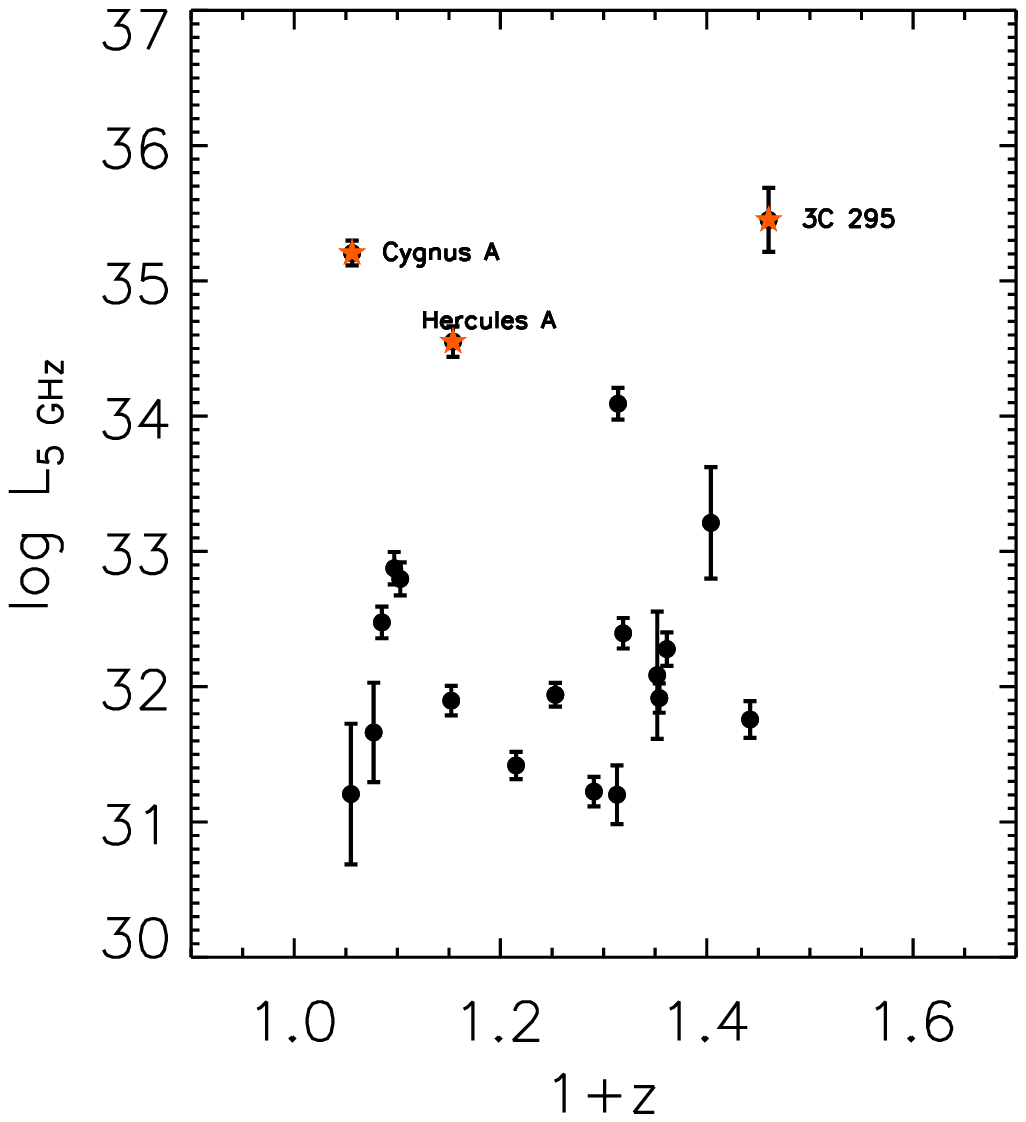}
\end{minipage}
\caption[Radio luminosities as a function of redshift]{Left - Logarithm of the total radio luminosity in $\ergps$ as a function of redshift over the 10 MHz to 10 GHz range. The Perseus cluster, Hercules A, H1821+643, Cygnus A and 3C 295 are highlighted with the stars. Right - Same but focused on the radio emission seen at 5 GHz. Also shown in the right panel are the errors bars, which have been multiplied by a factor of five for illustrative purposes. }
\label{fig5_5}
\end{figure*}

\section{Radio luminosities}
 
In this section, we focus on the radio properties, which can also be used as a tracer of black hole activity. Most of the radio emission in AGN arises from relativistic particles spiralling around magnetic fields and therefore emitting synchrotron emission. This type of emission is characterised by a steep radio spectrum with $\alpha>1$ ($S_\nu\propto\nu^{-\alpha}$). The radio lobes, observed to correlate with X-ray cavities in clusters are dominated by this emission. At higher radio frequencies ($\approx$GHz), emission from the core starts to dominate and is characterized by a flat or inverted spectrum with $\alpha\approx0$. Here, the core refers to the innermost brightest region of the jetted emission. We therefore consider more than one frequency in our analysis.

Nine of our clusters have already been analysed in detail by \citet{Bir2008686} at radio wavelengths. These authors derive flux density values at 327 MHz, 1.4 GHz, 4.5 GHz and 8.5 GHz. They also derive total radio luminosities by fitting a model to the spectral distribution of the source between 10 MHz and 10 GHz. These values are shown in Table \ref{tab5_4}. Note that, for some clusters, \citet{Bir2008686} derive both total and partial flux densities, the latter of which only includes the contribution of the radio lobes. We consider the total contributions. The radio properties of the MACS clusters are reported in HL2012. These values are also reported in Table \ref{tab5_4}. For the remaining clusters, we proceeded in the same manner as in HL2012, and searched for radio emission associated with the central AGNs by using various radio surveys publicly available. 

First, we searched for evidence of emission at 74 MHz with the $VLA$ Low frequency Sky Survey \citep[VLSS;][]{Coh2007134}, at 150 MHz with the TIFR GMRT Sky Survey (TGSS\footnote[2]{http://tgss.ncra.tifr.res.in}) and at 326 MHz with the Westerbork Northern Sky Survey\footnote[3]{http://www.astron.nl/wow/testcode.php?survey=1} \citep[WENSS;][]{Ren1997124}. Next, we searched for emission at 843 MHz using the Sydney University Molonglo Sky Survey \citep[SUMSS;][]{Boc1999117,Mau2003342}. We determined radio fluxes at 1.4 GHz using the 1.4 GHz $VLA$ Faint Images of the Radio Sky at Twenty-Centimeters survey \citep[FIRST;][]{Bec199461}. If FIRST data were not available, we derive the fluxes using the 1.4 GHz NRAO $VLA$ Sky Survey catalogue \citep[NVSS;][]{Con1998115}. Finally, we searched for high frequency radio emission associated with the central AGNs using the 5 GHz Parkes-MIT-NRAO (PMN) radio survey \citep{Gri1993105} and the compilation of results from EINSTEIN Observatory Extended Medium-Sensitivity Survey \citep[EMSS;][]{Gio199072,Sto199176}. Finally, at 28.5 GHz, we used the Berkeley-Illinois-Maryland Association survey \citep[BIMA;][]{Cob2008134}. As in HL2012, if no point source was seen within 100 kpc of the BCG, we used the 2$\sigma_{\rm rms}$ value within the beam area as an upper limit to the flux.

For some of the radio surveys, the errors quoted in the catalogue did not account for systematic uncertainties. Systematic uncertainties vary with frequency, but are on the order of 5 per cent \citep[][]{Car1991383}. If a survey did not include systematic uncertainties, we computed the total uncertainty assuming a 5 per cent systematic error and a 2$\sigma_{\rm rms}$ noise level. In this case, errors are derived as the quadratic sum of the rms noise level in the map and the systematic uncertainty associated with the value.  

Finally, we computed a rough estimate of the total radio luminosity between $\nu_1=10$ MHz and $\nu_2=10000$ MHz following Eq. \ref{eq5_2}, where $D_{\rm L}$ is the luminosity distance to the source and $S_\nu$ is the flux density. 
\begin{eqnarray}
L_{\rm radio}=4{\pi}D^2_{\rm L}\int_{\nu_1}^{\nu_2}(S_\nu)d\nu
\label{eq5_2}
\end{eqnarray}
The flux densities at 10 MHz and 10 GHz are extrapolated from the other known values based on the assumption that the flux density scales as $S_\nu\propto\nu^{-\alpha}$. Here, the local spectral index determined from the two nearest flux density data points are used to compute the extrapolated values. Using a simple trapezoid rule, we then integrated over the 10 MHz to 10 GHz range. For MACS J0242.5$-$2132, MACS J1411.3+5212, PKS 0745-19, 4C+55.16, A2029 and Hercules A, we also used the extensive follow-up radio observations available in the NASA/IPAC Extragalactic Database (NED) to determine $L_{\rm radio}$. Our results are shown in Fig. \ref{fig5_5}. Since high frequency radio observations are better proxies of the core emission, we also show in Fig. \ref{fig5_5} the 5GHz radio luminosities of our sources as a function of redshift, but stress that the scatter in the radio figures is larger than that observed at X-ray wavelengths (Fig. \ref{fig5_4c}). We therefore concentrate only on the X-ray evolution of our sources in the following sections. 
\begin{figure}
\centering
\begin{minipage}[c]{0.9\linewidth}
\hspace{-1cm}
\centering \includegraphics[width=\linewidth]{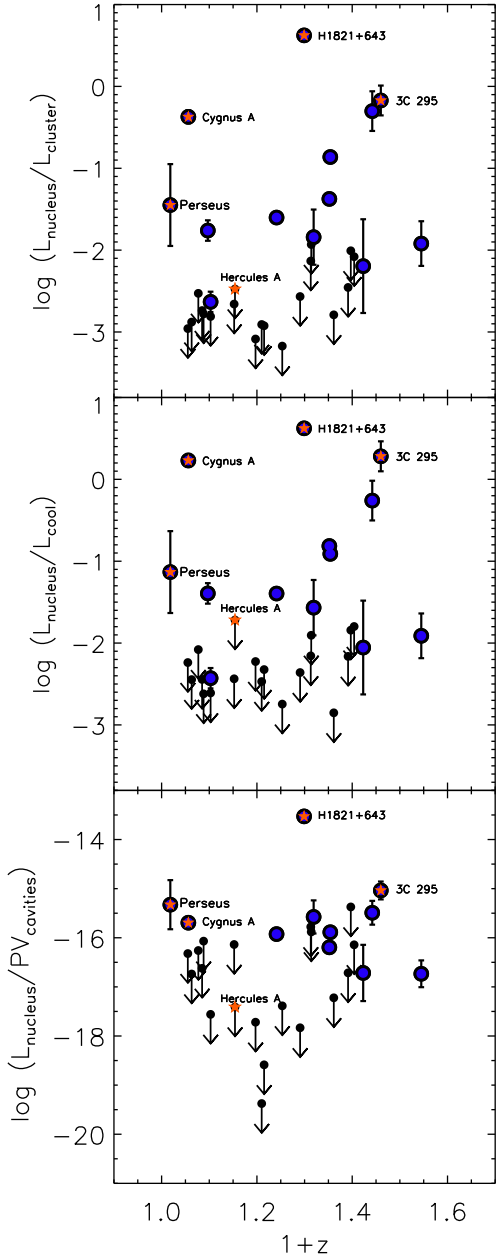}
\end{minipage}
\caption[Nuclear X-ray luminosities corrected for the cooling and cluster luminosities, as a function of redshift]{Ratio of the nuclear X-ray luminosities corrected for total cluster and cooling luminosities, as well as cavity energy ($PV_{\rm cavities}$). }
\label{fig5_6}
\end{figure}

\section{Results and selection effects}

Fig. \ref{fig5_4c} shows that the nuclear X-ray emission of BCGs is evolving with time such that they are on average brighter in the past. These BCGs have been selected such that they have clear AGN-driven X-ray cavities, and all lie in cool-core clusters. They therefore represent a subset of the BCG population and the evolution seen may not necessarily apply to all BCGs. In other words, we only consider those with strong radio-mode feedback taking place, and trace how the radiative evolution of these systems evolves with time. 

We stress that there is no clear selection bias against the lower right portion of Fig. \ref{fig5_4c}, i.e. the portion where high-redshift BCGs have low nuclear X-ray luminosities. Indeed, when selecting our sample, our main criteria is whether a massive cluster has X-ray cavities. The detectability of X-ray cavities does not depend on the central AGN being switched on or not radiatively, since $Chandra$ can resolve the nuclear point source from the cavities. If the nuclear X-ray emission of 3C 295 was 100 times fainter, we still would have included it in our sample since we still would have seen the X-ray cavities. Yet, the lower right portion of Fig. \ref{fig5_4c} remains unpopulated, pointing to some form of evolution of the nuclear properties in BCGs with X-ray cavities. In this section, we investigate other possible selections effects that may alter our results.

First, we discuss BCG duty cycles in cool-core clusters. Studies of nearby clusters show that almost all cool-core clusters have X-ray cavities \citep[detection rates $>90$ per cent, see][]{Dun2006373,Fab2012}, suggesting that BCG-hosted AGN duty cycles are large ($>90$ per cent) in cool-core clusters, at least in terms of mechanical outflows. A possible selection effect could be that we are missing some clusters that currently have no apparent X-ray cavities and therefore have not been included in our sample due to our selection criteria, but these represent at most 10 per cent of the cool-core population and should not affect our results significantly if the duty cycles remain high at high redshifts. 

An important selection effect is that we could be missing systems with X-ray cavities at high redshift, not due to the AGN being switched off mechanically, but rather due to the increasing difficultly in detecting X-ray cavities at high redshift combined with a lack of deep data for these objects. In HL2012, we found that at least 19 out of 37 MACS clusters with cool cores (50 per cent) seemed to host cavities. These 19 clusters included 13 with clear cavities as well as 6 with less well defined cavities and only classified as potential cavities (see HL2012 for more details). More importantly, the majority of the remaining 18 cool-core clusters with no detectable X-ray cavities have less than 30 ks $Chandra$ observations. A possible selection effect could therefore be that we are missing a significant population of MACS clusters with X-ray cavities for which the data quality is not sufficient to detect cavities but deeper data for these objects would be needed. While being mindful of this possible effect, for now, we base our results on the current population of BCGs with known X-ray cavities.  

Third, it is possible that we are missing luminous clusters at high redshift not initially included in MACS due to the flux limited nature of the survey. New surveys based on the Sunyaev-Zel'dovich effect which has no surface brightness dimming, are starting to come online. These include those based on the South Pole Telescope \citep[SPT][]{Car2011123,Van2010722} and $Planck$ (Bartlett et al. 2009, Planck Collaboration 2011a)\nocite{Bar2008329,Planck2011}, and will provide a wealth of new data in the intermediate to high redshift Universe.

Note that, as we are increasing in redshift, we are probing a larger comoving volume. A possible selection effect could be that if clusters with radiatively efficient central AGN are rare, as we increase in redshift, the probability of encountering one increases. This could explain why we detect such objects at high redshift, but can not explain the lack of very faint AGN at high redshifts which should remain just as common if there was no evolution. The evolution we find is also steeper than the expected increase in comoving volume with redshift for our cosmology, based on the \citet{Row1968138} test \citep[see also][]{Row2010406}, so this effect cannot explain entirely our results.  

Another possible selection effect could be that we are selecting only the brightest clusters at high redshift, since these are the ones where X-ray cavities are more easily detected. However, the top panel of Fig. \ref{fig5_6} shows that even if we correct the cluster X-ray luminosity, we still see that higher-redshift BCGs are more X-ray active than their low-redshift counterparts. Here, we use the cluster luminosities shown in column 4 of Table \ref{tab5_2}. In this case, a linear regression in the log-log space is consistent with a positive slope at a 92 per cent level.

We also investigate the possibility that the evolution seen is not due to the intrinsic evolution of the nuclear luminosities, but instead caused by the evolving properties of the cluster. For lower redshift clusters at $z<0.5$, both HL2012 and \citet{Bau2005359} show that the fraction of cool-core clusters to non cool-core clusters remains fairly constant up to present day, with almost half of the clusters showing evidence of  short central cooling times. No evolution in the cool-core properties of clusters are seen, at least up to $z=0.5$ (however, see Samuele et al. 2011 and McDonald 2011 for evidence of evolution in terms of the optical properties of BCGs)\nocite{Sam2011731,McD2011742}. At $z>0.5$, several authors have shown that their appears to be a significant lack of $very$ strong cool-core clusters \citep{Vik2007,San2008483,San2010521,All201149}, indicating that there may be some form of rapid evolution in the cool-core properties of clusters between $z=0.5$ and $z=1$. However, \citet{Sem2012} recently identified $z>0.5$ strong cool core clusters based on the SPT survey. The observed lack of $very$ strong cool-core clusters at $z>0.5$ could therefore simply be due to an identification bias (see also Section 8.3). Our sample consists of clusters within $z=0.6$. At least in terms of the cool core properties of the clusters, there should be no significant evolution, and this should not explain the trend we see in Fig. \ref{fig5_4c}. To illustrate this, we have computed the cooling luminosities of all of our clusters in Table \ref{tab5_2} and show in the middle panel of Fig. \ref{fig5_6} the nuclear X-ray luminosities corrected for cooling luminosity. The linear regression remains consistent with a positive slope to a 90 per cent.

\begin{figure}
\centering
\begin{minipage}[c]{0.99\linewidth}
\centering \includegraphics[width=\linewidth]{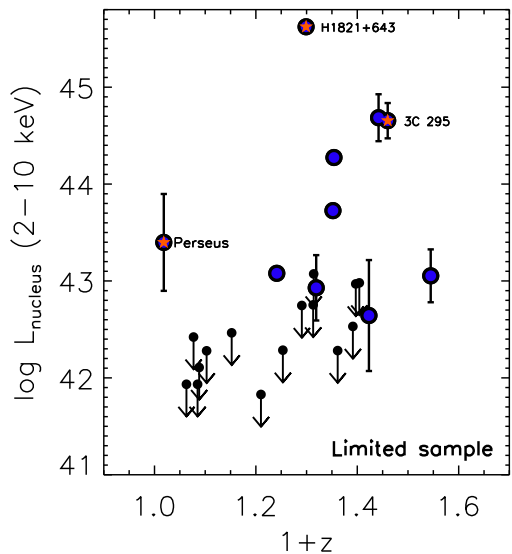}
\end{minipage}
\caption[]{Same as Fig. \ref{fig5_4c}, but we limit our sample to 26 clusters instead of 32. We remove the weakest cool core clusters at $z<0.3$ (Abell 0085, Cygnus A, RXC J1524.3-3154, RXC J1558.3-1410, Hercules A, Abell 0115 and ZwCl 2701) such that the average cooling luminosity for our $z<0.3$ clusters is the same as the one for our $z>0.3$ clusters. In other words, we further limit the sample so that we are looking at the same population of cool core clusters with X-ray cavities across all redshifts. }
\label{fig5_6a}
\end{figure}

According to Table \ref{tab5_2}, our $z>0.3$ clusters have cooling luminosities ($L_{\rm cool}$) 2 to 3 times higher than our $z<0.3$ clusters. Larger cooling luminosities implies the need for stronger feedback from the central AGN to counterbalance the cooling, and may therefore result in a higher accretion rate and brighter AGN. The observed evolution in Fig. \ref{fig5_4c} could therefore simply be due to a selection bias where we selected on average weaker cool cores at low redshift. We test this theory in Fig. \ref{fig5_6a}, where we limit the sample to 26 clusters instead of 32 such that the $z>0.3$ population of clusters has on average the same cooling luminosity as the $z<0.3$ population. To do this, we simply remove the weakest cool core clusters until the average cooling luminosities are within 10 per cent of each other. This results in the removal of Abell 0085, Cygnus A, RXC J1524.3-3154, RXC J1558.3-1410, Hercules A, Abell 0115 and ZwCl 2701, all of which are the weakest cool core clusters in our sample, i.e. the clusters with the smallest cooling luminosities. If we limit the sample to the remaining 26 clusters, and plot the $2-10\keV$ nuclear luminosity as a function of redshift for this limited sample, we see from Fig. \ref{fig5_6a} that the apparent evolution is even more pronounced. Here, the slope of a linear regression is positive to more than a 96 per cent level, indicating that the evolution truly seems to be real.

Finally, we analyze in the bottom panel of Fig. \ref{fig5_6} the nuclear X-ray luminosities corrected for cavity energy, where we estimate that the energy stored within each cavity is given as the multiplication between the thermal pressure of the surrounding hot gas and the cavity volume ($PV_{\rm cavities}$). The cavity energetics shown in Fig. \ref{fig5_6} were taken from the literature \cite[see Table \ref{tab5_2};][and HL2012]{Bir2004607,Dun2005364,Raf2006652,Dun2006373,Dun2008385,Hla2011415}, but RXC J1524.3$-$3154 or RXC J1558.3$-$1410 have not been included in the plot since the cavity energetics are not yet available. For the objects where multiple cycles of outbursts can be seen (e.g. Perseus), we only consider the inner X-ray cavities. Fig. \ref{fig5_6} shows that when correcting for cavity energy, the scatter is larger but the slope remains positive, indicating that the evolution seen is mostly due to the evolution of the radiative properties of the BCGs. We further discuss this plot in Section 8.1.3. 

Note that, we have used the cavity enthalpy as a proxy of the outflow energetics, and not the cavity power ($P_{\rm cavity}=4PV_{\rm cavities}/t_{\rm age}$) as the literature often shows different estimates of a cavity age (e.g. sound crossing time and/or buoyancy rise time), and it remains unclear which one is the most accurate. To avoid the added scatter of cavity age, we have chosen to illustrate Fig. \ref{fig5_6} using simply the enthalpy ($PV_{\rm cavities}$). Note also, Perseus is the only source that clearly shows variability of the nuclear emission over yearly time scales, and we include an error bar for this source that takes this variability into account. It is possible that other BCGs also exhibit such variability, but it is not clear if this is the case for all BCGs. 4C+55.16 and PKS0745-19 are two example that show no significant variation on 5-10 year time scales, see \citet{Hla2011415} and Sanders et al. in preparation. 

In summary, selection effects are important, and while being mindful of them, we base our discussion in the following sections only on the currently known population of BCGs that lie in massive clusters of galaxies with known X-ray cavities.

\begin{figure*}
\centering
\begin{minipage}[c]{0.6\linewidth}
\hspace{-1cm}
\centering \includegraphics[width=\linewidth]{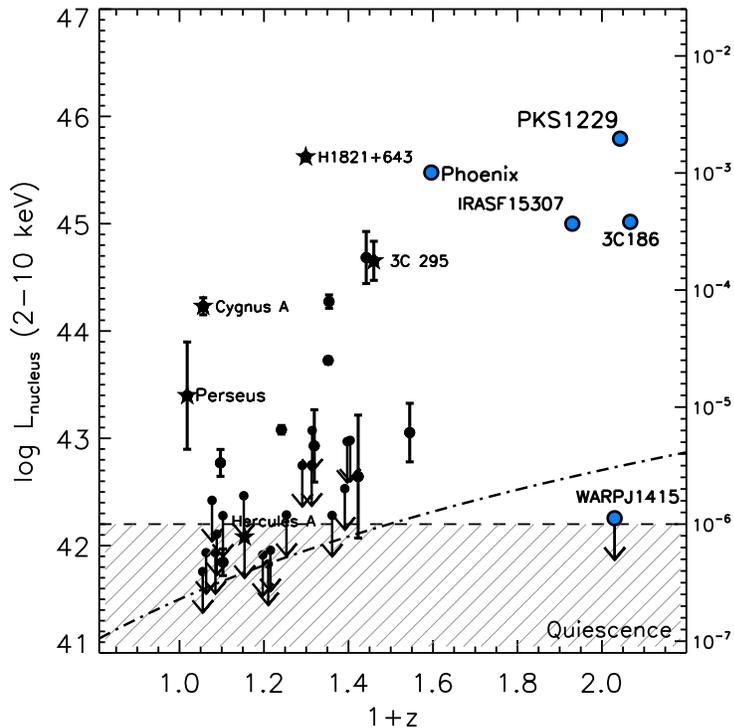}
\end{minipage}
\caption[Nuclear X-ray luminosities as a function of redshift]{Nuclear X-ray luminosities as a function of redshift. The black filled circles illustrate the data points from Fig. \ref{fig5_4c}. The downward pointing arrows represent upper limits to the nuclear luminosities. The high-redshift sources discussed in Section 7 are shown with the blue points. The black dashed curve shows observed evolution seen in star formation efficiency ($L_\star\propto(1+z)^{4}$). The right Y-axis shows the Eddington ratio of the nuclear X-ray luminosity assuming a $10^{10}{\rm M_\odot}$ black hole and the striped region shows the location where a black hole reaches quiescence ($L_{\rm nucleus}\aplt10^{-6}{\rm L_{\rm Edd}}$). }
\label{fig5_8}
\end{figure*}

\section{High-redshift clusters of galaxies}
There are a number of recent studies that have analysed individual $z\approx1$ clusters, and found evidence for both AGN activity and short central cooling times. These include WARP J1415.1+3612 at $z=1.03$ \citep{San2012539}, XMMUJ0044.0-2033 at $z=1.579$ \citep{San2011531}, PKS 1229-021 at $z=1.04$ \citep{Rus2012422} and 3C186 at $z=1.067$ \citep{Sie2010722}. Although none of these studies have provided clear evidence for the existence of X-ray cavities being carved out by the radio lobes, most likely due to the limited resolution, the X-ray luminosities of these clusters are on the order of $10^{44}\ergps$ and all of these show at least tentative evidence of a cool core. These clusters are therefore similar to our sample of objects, and if BCG-hosted AGN duty cycles remain high in high-redshift cool-core clusters, most should harbour X-ray cavities. We therefore plot these in Fig. \ref{fig5_8}. Note that we did not include these in our original sample since they do not show evidence of X-ray cavities. Fig. \ref{fig5_8} does not show XMMUJ0044.0-2033 since \citet{San2011531} do not derive flux estimates for any non-thermal contribution of a central AGN, i.e. they only focus on deriving fluxes for the thermal cluster emission. There are also no $Chandra$ observations available for this source in order to isolate the nuclear emission of the BCG.

As for WARP J1415.1+3612, \citet{San2012539} analysed very deep $Chandra$ observations of this source ($278\ks$), but did not analyse the X-ray emission from the central AGN. We have reprocessed the $Chandra$ observations (Obs ID 12255, 12256, 13118 and 13119), and searched for evidence of non-thermal emission associated with the central AGN using the method outlined in Section 4.2. In this case, we considered a region centred on the BCG within a radius of 1$''$ as our nuclear region and a surrounding annulus located within $r=2''-3''$ as our background. Applying C-statistics to account for the low number of counts, we find that both an absorbed (Galactic) power-law and {\sc mekal} model provide an equally good fit to the data. There is no clear evidence from the spectrum that the source contains non-thermal emission. Fitting a combined {\sc mekal + power-law} model to the data does not improve the fit, and even when freezing most parameters, the fit remains uncontrained. We therefore derive an upper limit to the non-thermal emission using the same method outlined in Section 4.1 where we convert a count rate into a flux. Throughout the analysis, we have assumed the same galactic absorption and central location for our regions as in \citet[][; $N_{\rm h}=1.05\times10^{20}{\rm cm^2}$, RA=14:15:11.08 and DEC=+36:12:03.1]{San2012539}. We derive a $3\sigma$ upper limit of $1.8\times10^{42}\ergps$ for the nuclear emission of WARP J1415.1+3612. This result is shown in Fig. \ref{fig5_8}.

Fig. \ref{fig5_8} also includes the powerful quasar and hyper-luminous infrared galaxy IRAS F15307+3252 \citep[][]{Iwa2005362,Fab1996283}, which is a Compton-thick AGN embedded in a luminous cluster ($L_{\rm X-ray, cluster}\approx10^{44}\ergps$). The estimated $2-10\keV$ luminosity is on the order of $10^{45}\ergps$. Although there is yet no known evidence indicating the presence of a cool core or X-ray cavities, we include it in Fig. \ref{fig5_8} for comparison. Finally, we include the $z=0.596$ extreme cluster SPT-CLJ2344-4243 in Fig. \ref{fig5_8}, also known as the Phoenix cluster. This cluster is highly X-ray luminous ($L_{\rm X-ray, cluster}\approx10^{45-46}\ergps$), harbours a strong cool core as well as a massive starburst at its core \citep[$740{\rm M_{\rm \odot}}{\rm yr^{-1}}$;][]{McD2012Nat}. No X-ray cavities have yet been reported in this object and we therefore only include it in Fig. \ref{fig5_8} for illustrative purposes.

Interestingly, Fig. \ref{fig5_8} shows that if we were to extrapolate the apparent evolution seen in our sample of BCG-hosted AGN out to $z=1$, some of the currently known high-redshift clusters with evidence of AGN activity and short central cooling times lie within this extrapolation. At least in terms of the radiatively efficient high-redshift sources, these objects may be the less evolved counterparts of nearby massive cool-core clusters with powerful AGN-driven outflows. Note that WARP J1415.1+3612 may simply be absorbed and therefore undetectable at X-ray wavelengths with $Chandra$. It therefore does not necessarily represent a distinct population of objects.

\section{Discussion}

\subsection{Implications for the transition between quasar-mode and radio-mode feedback}

\begin{figure*}
\centering
\hspace{-1cm}
\begin{minipage}[c]{0.38\linewidth}
\centering \includegraphics[width=\linewidth]{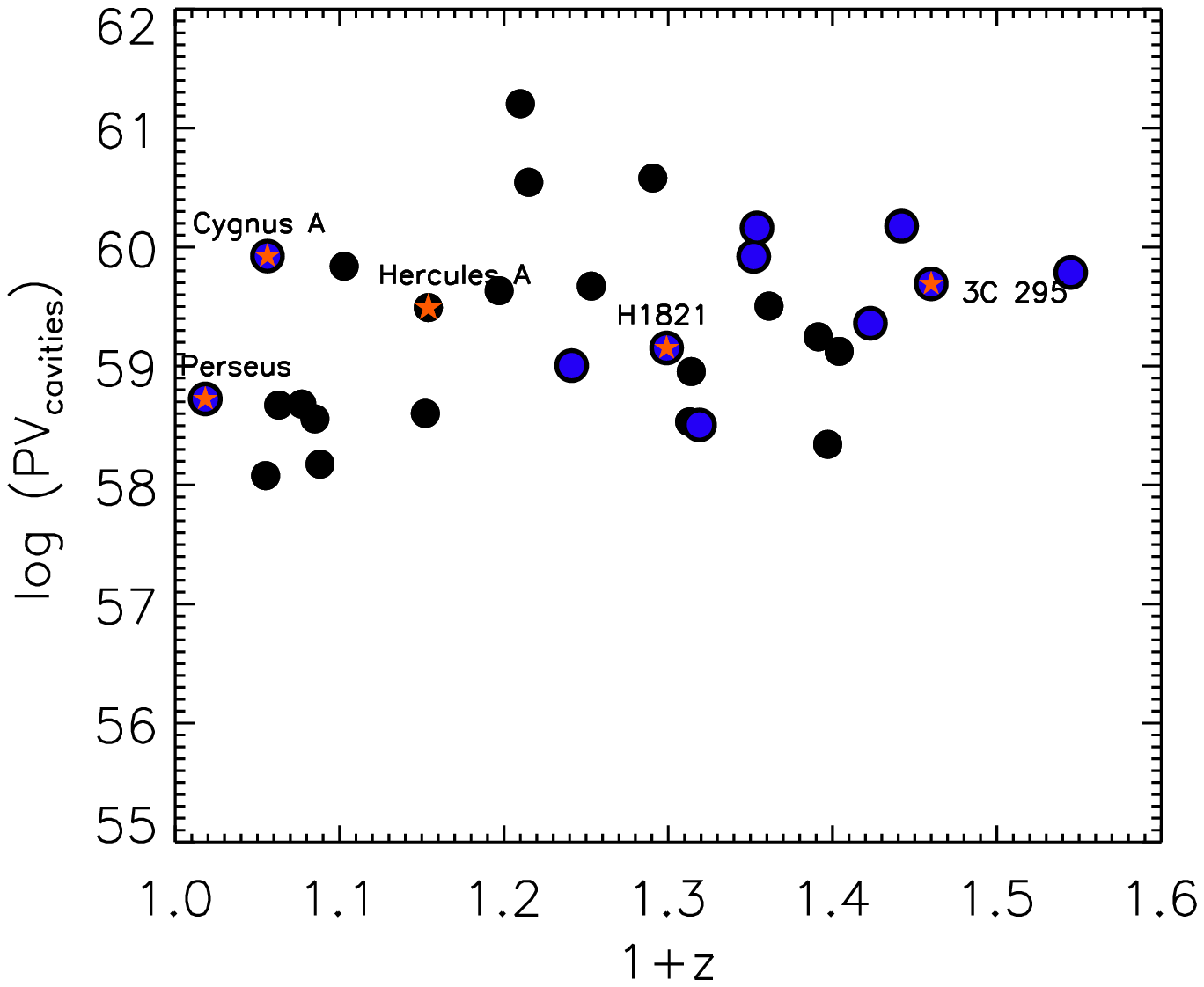}
\end{minipage}
\hspace{-0.6cm}
\begin{minipage}[c]{0.33\linewidth}
\centering \includegraphics[width=\linewidth]{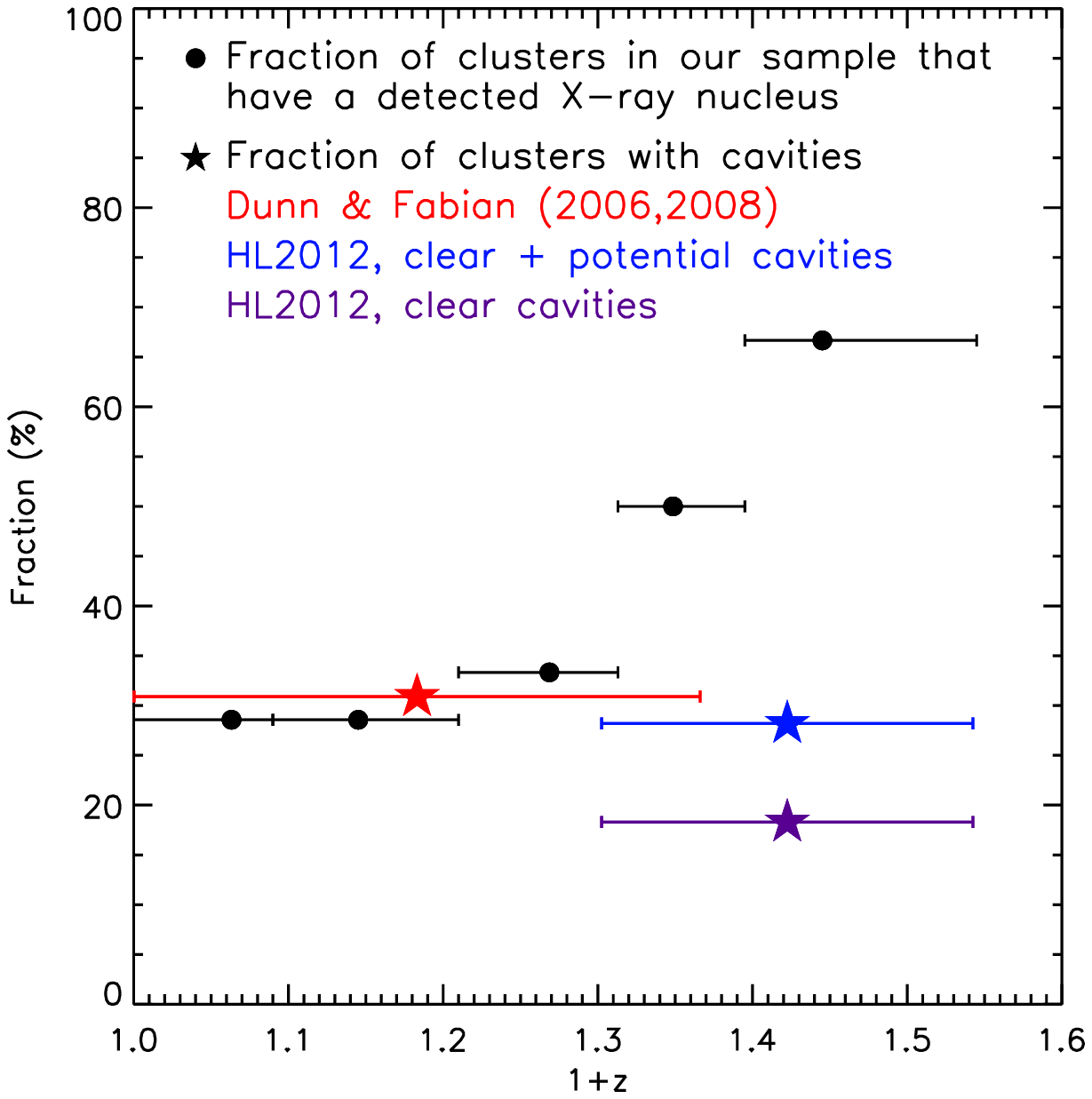}
\end{minipage}
\hspace{-0.5cm}
\begin{minipage}[c]{0.33\linewidth}
\centering \includegraphics[width=\linewidth]{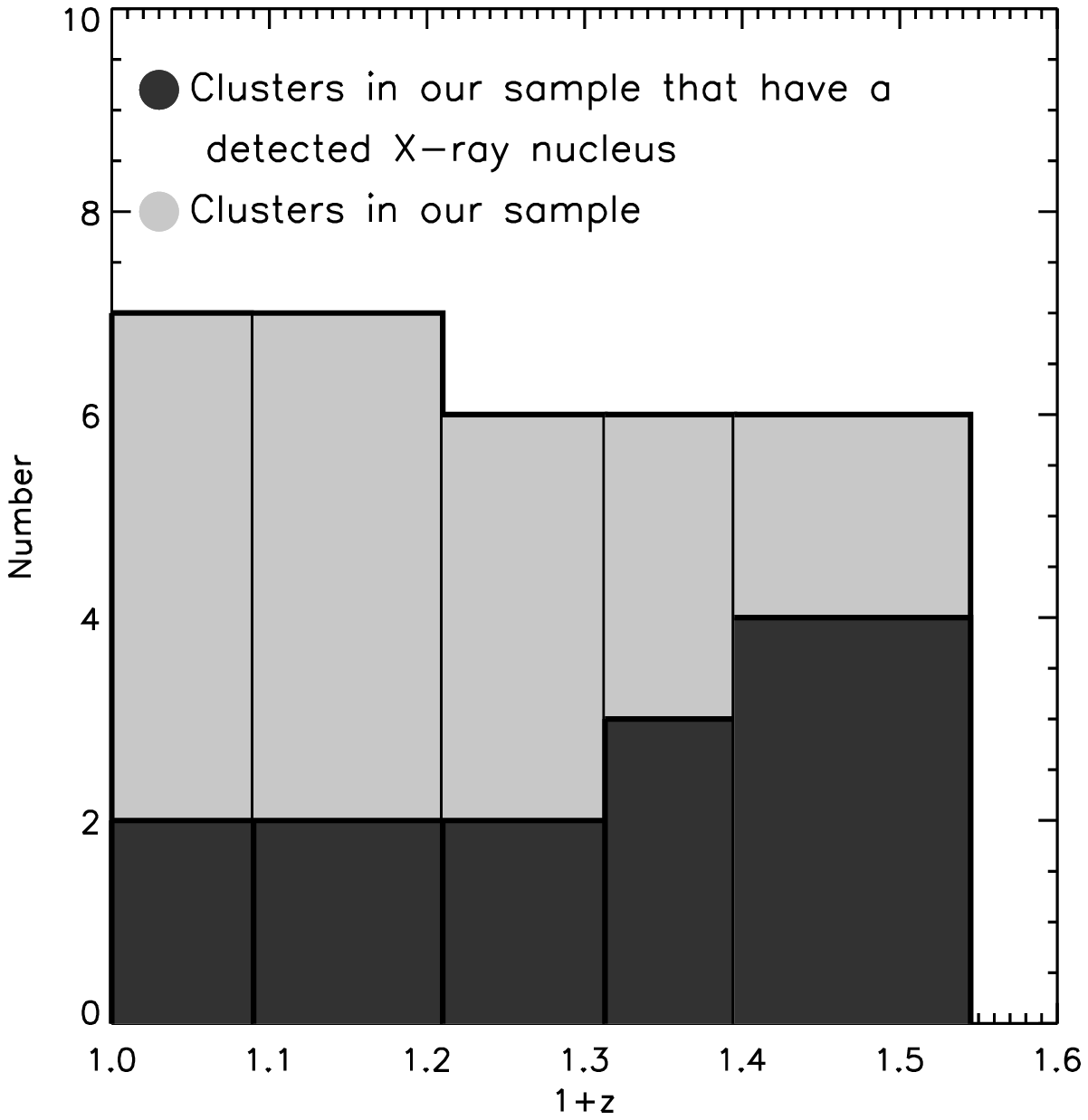}
\end{minipage}
\caption[]{\textit{Left}: Cavity enthalpy as a function of redshift for the sources in our sample (see Table \ref{tab5_2}). \textit{Middle}: Fraction of BCGs in our sample with a central detected X-ray nucleus as a function of redshift (black). Also shown are the fraction of galaxy clusters with X-ray cavities based on the low-redshift sample of \citet[][; $\approx30$ per cent]{Dun2006373,Dun2008385} and the $z>0.3$ sample of HL2012 ($\approx20-30$ per cent). For the latter, we show the fraction in terms of the clear and potential cavities (light blue), as well as the fraction in terms of only the clear cavities (purple). \textit{Right}: Histogram of the number of sources in our sample with a detected X-ray nucleus. Note that, our sample only includes BCGs that lie in massive clusters and that have known X-ray cavities.}
\label{fig5_9}
\end{figure*}

In HL2012, we analysed the properties of X-ray cavities in the MACS sample and compared these to low-redshift sample of \citet{Raf2006652}. Interestingly, we found no significant evolution of the X-ray cavity properties with redshift. We illustrate this in terms of the cavity energetics in the left panel of Fig. \ref{fig5_9}, where we plot the cavity enthalpy as a function of redshift for the sample studied here. Fig. \ref{fig5_9} shows that there is no significant evolution in the cavity energetics (less than an order of magnitude), at least in terms of the cavities found in massive cool core clusters. 

On the other hand, Fig. \ref{fig5_8} shows that the sample studied here appears to be rapidly evolving in terms of the nuclear X-ray luminosities, suggesting that we are seeing some form of evolution of how AGN feedback operates in BCGs with X-ray cavities over the $z=0-0.6$ range. Such an evolution was first noted by \citet{Hal1997113}, who analyzed a small subset of 3 quasar-like BCGs and found evidence for evolution. Here, we consider a significantly larger sample of BCGs and analyze the implications of our results. We investigate two possible scenarios that could explain our results, and dicuss them in the following two sections.

\subsubsection{Evolution in terms of the general increase in luminosity with redshift}

\begin{figure*}
\centering
\begin{minipage}[c]{0.6\linewidth}
\hspace{2cm}
\centering \includegraphics[width=\linewidth]{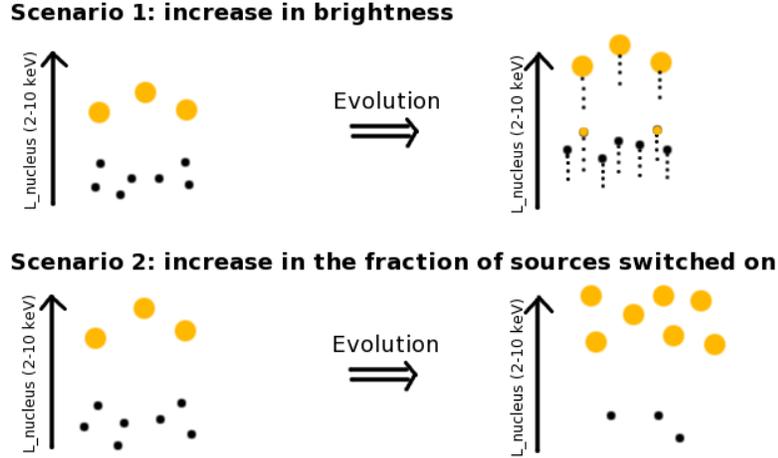}
\end{minipage}
\caption[]{Two possible scenarios to explain the evolution of the nuclear X-ray luminosities in our sample of BCGs with X-ray cavities as a function of redshift (redshift increases from left to right). The orange circles illustrate AGN that have a detectable X-ray nucleus, whereas the black colors illustrate the BCGs with no detectable X-ray nucleus. The top panel shows the first evolution scenario, such that BCGs with X-ray cavities become brighter with increasing redshift (from left to right). As they become brighter, they become more easily detected, hence the increase in the number of orange points. The bottom panel shows the second scenario, in which the fraction of radiatively efficient sources increases with redshift, from 30 per cent at $z\approx0.1$ (left) to some 60-70 per cent at $z\approx0.6$ (right).  }
\label{fig5_10}
\end{figure*}

We first consider the possibility that we are seeing a general increase in the nuclear X-ray luminosity of BCGs with increasing redshift (see top panel of Fig. \ref{fig5_10}), in the sense that BCG-hosted black holes are rapidly evolving such that their $2-10\keV$ X-ray luminosity has increased with redshift by an order of magnitude over the last 5 Gyrs.

Our sample of BCGs only includes those in the most massive clusters of galaxies and that have clear AGN-driven outflows in the form of X-ray cavities. We are therefore only considering a subset of the BCG population. We also stress that we are only probing the $0.5-7\keV$ energy range as seen from $Chandra$. It is not clear if the evolution seen can be applied to all wavelengths. 

Nevertheless, the suggestive evolution seen in Fig. \ref{fig5_4c} is consistent with being steep. In comparison, the star formation efficiency in galaxies evolves approximately as $L_\star\propto(1+z)^{4}$ since $z=1$ \citep[e.g.][]{Ruj2010718,Lil1996460,Mad1996283,Hop2004615}. A similar evolution is seen in terms of the number density of quasars which peaks at a redshift of $z\approx2$ and then rapidly declines to the present day \citep{Ued2003598,Has2005441,Hop2007654,Air2010401}. The exact shape of density function depends on the luminosity bin considered (higher luminosity AGN peak at higher redshift), as well as the wavelength considered, but the decline from $z\approx1$ to the present day follows roughly a $(1+z)^{4}$ scaling at all wavelengths \citep[e.g. ][]{Hop2007654}. Fig. \ref{fig5_8} highlights our data points as well as the typically observed $(1+z)^{4}$ scaling with the dot-dashed line. If we calculate the average nuclear luminosity for our $0.3<z<0.6$ BCGs, we find that it is at least 10 times higher than that of the BCGs in the $0<z<0.3$ redshift range, yet the star formation efficiency increases only by a factor of $2-3$ over these redshift bins. BCGs would therefore appear to be changing more rapidly than the general population of AGNs and the decline of star formation activity in galaxies. It is possible that the unique environments of BCGs, and especially those in our sample which lie in the most extreme cool-core clusters, cause their central AGN to shut down radiatively more rapidly than others, perhaps due to higher merger rates at the centres of clusters.

Quasar-mode feedback and its evolution have also been implemented in various simulation works \citep{DiM2005433,Cro2006365,Bow2006370,Spr2005435,Sij2006366,Mer2008388,Dub2012420}.
\citet{Cro2006365} and \citet{Mer2008388} predict that the black-hole accretion rate density (related to luminosity density assuming a certain efficiency) for black holes operating in quasar-mode feedback declines at most by a factor of ten since $z\approx2$. Their simulations also predict that the accretion rate density for black holes operating in radio-mode feedback (or kinetic-mode feedback) remains roughly the same since $z\approx2$. Although both these pictures agree well with our BCGs in terms of radio-mode feedback, since we find no evidence for evolution in the outflow mechanical properties in HL2012, the evolution of quasar-mode feedback may be significantly steeper for BCGs with X-ray cavities than predicted by these simulations, i.e. a factor of 10 since $z\approx0.5$ as opposed to a factor of 10 since $z\approx2$. 

The right hand axis of Fig. \ref{fig5_8} shows the nuclear $2-10\keV$ luminosity of our sources in terms of the Eddington ratio. To compute the Eddington ratio, we have assumed a typical black hole mass of 10$^{10}{\rm M_\odot}$ for our sample of BCGs since they lie in the most luminous clusters of galaxies, i.e. BCG mass and therefore SMBH mass scales with cluster X-ray luminosity. Thus, it is reasonable to assume that these BCGs will host some of the most massive black holes with regards to other BCGs and will have typical average masses of $10^{10}{\rm M_\odot}$ \citep[see also][]{McC2011480,Hla2012}. There is also evidence suggesting that the SMBHs in BCGs may not follow the standard $M_{\rm{BH}}-\sigma$ or $M_{\rm BH}-M_{K}$ correlations observed in massive galaxies (see von der Linden et al. 2007, Dalla Bont\`a et al. 2009, Lauer et al. 2007, Hlavacek-Larrondo et al. 2012)\nocite{Lau2007662,von2007379,Dal2009690,Hla2012}. For the purposes of this study, we therefore simply assume that our black holes have $10^{10}{\rm M_\odot}$ masses. 

In terms of the Eddington ratios, Fig. \ref{fig5_8} shows that the evolution is steep, from $\approx10^{-3}{\rm L_{\rm Edd}}$ at $z=1$ to $\approx10^{-6}{\rm L_{\rm Edd}}$ at $z=0.1$. \citet{Dub2012420} traced the growth of black holes using cosmological simulations, while implementing both radio-mode feedback, in the form of a jet-like outflow when the black hole is accreting at low rates, and quasar-mode feedback, in the form of heat when the black hole is accreting at high rates. This allowed them to trace how black holes transit from one mode to the other. In terms of the evolution of the Eddington ratios, their black holes evolve on average from a ratio of $\approx10^{-3}{\rm L_{\rm Edd}}$ at $z=1$ to $\approx10^{-4}{\rm L_{\rm Edd}}$ at $z=0$. Interestingly, our results agree with \citet{Dub2012420} at high redshift ($z=1$). However, in terms of the evolution, our results show that BCGs with X-ray cavities evolve much faster (about 100 times faster) from quasar-mode to radio-mode feedback than predicted by these simulations for the general population of black holes, although for the more massive black holes the simulations show that the evolution might be larger \citep[see Fig. 14 in][]{Dub2012420}. 

\subsubsection{Evolution in terms of the number of sources switched-on radiatively}

We also investigate a second possibility as to why BCGs with X-ray cavities appear to be more radiatively efficient at high redshifts. Our sample only contains a small number of sources, especially at high redshifts, and could therefore be affected by small number statistics as well as selection effects as seen in Section 6. Yet, we find that the high redshift sources are brighter, and that the majority have a detectable X-ray nucleus. 

If there was no evolution and assuming that we have not been biased towards only radiatively efficient sources at high redshift, we should have detected the same fraction of sources with bright, radiatively efficient nuclei and therefore easily detectable at all redshifts. Yet, we find that the majority of high redshift sources have a bright and detectable central AGN, whereas the majority of the low redshift ones, even with very deep observations, have no detectable central X-ray nucleus. This points to evolution, and we investigate the possibility that the evolution seen is simply be due to the increasing fraction of sources switched-on radiatively, as opposed to a general increase in brightness for the BCG population with X-ray cavities. In other words, there would be two states, one radiatively efficient and one inefficient and as time progresses, BCGs would jump from the efficient to the inefficient state, making them largely undetectable at X-ray wavelengths. 

We illustrate this second possibility in the middle and right panels of Fig. \ref{fig5_9} where we plot the fraction of BCGs in our sample that have a detectable X-ray nucleus as a function of redshift. Here, we have chosen the redshift bins such that each contains 6 or 7 BCGs in them. Since our sample only contains 32 sources, the plot is affected by small number statistics and varies depending on our definition of the redshift bins. However, in all cases, we see that the fraction roughly doubles over the last 5 Gyrs, and on average varies from 30 per cent at $z\approx0.1$ to 60 per cent at $z\approx0.6$. This result remains even when we consider the limited sample of Fig. \ref{fig5_6a} which only considers the 26 clusters with similar cooling luminosities instead of the original 32 clusters in our sample. 

The middle panel of Fig. \ref{fig5_9} also shows the fraction of clusters with X-ray cavities in two redshift bins (see the coloured points), highlighting the lack of strong evolution in the fraction of clusters with cavities. Here, we use the sample of 71 clusters in \citet[][]{Dun2008385}, which consist of a complete sample of cool-core and non cool-core clusters, to illustrate the average fraction of clusters with cavities at low redshift (red point), and compare this to the MACS clusters at high redshift (blue points). This figure therefore shows that the number of clusters with X-ray cavities remains roughly the same, yet those with cavities and detectable X-ray nuclei varies strongly with redshift. A possible scenario could therefore be that we are not seeing a gradual increase in X-ray luminosity of all of our sources, but rather an increase in the fraction of sources that are in a radiatively efficient state (see lower panel of Fig. \ref{fig5_10}).

\subsubsection{Brief note on radiatively efficient BCGs with X-ray cavities}

The bottom panel of Fig. \ref{fig5_6} illustrates the ratio between the nuclear X-ray luminosity and mechanical energy of the AGN-driven outflows in our sample as a function of redshift. An interesting result is that if we only consider the sources with detected AGN in our sample, the ratio between the radiative and mechanical output seems to remain fairly constant with redshift, apart for the only quasar in our sample, H1821+643. This might be indicating that once the nucleus is switched on radiatively, for instance once it reaches a radiatively efficient state, the power that emerges in the form of mechanical jets scales roughly proportionally with the radiative power (i.e. how much mass the black hole is accreting), regardless of redshift. This is discussed in more detail in \citet{Rus2013}. This result might be indicating a fundamental property of accretion physics. 

\subsection{Implications for the accretion modes in BCG-hosted black holes: analogy with black-hole binaries}

We now discuss the possibility that we are seeing a state transition of BCG-hosted black holes, operating from a more radiatively efficient state at high redshift to a radiatively inefficient state at low redshift \citep[see also][]{Chu2005363}. This interpretation can be applied to both scenarios proposed in Fig. \ref{fig5_10}. 

We recall that the right hand axis of Fig. \ref{fig5_8} shows the $2-10\keV$ luminosity in terms of the Eddington ratio, assuming a typical black-hole mass of 10$^{10}{\rm M_\odot}$ for our sample of BCGs. This figure shows that low-redshift BCGs with X-ray cavities are extremely radiatively inefficient and have typical Eddington ratios of $10^{-7}-10^{-5}{\rm L_{\rm Edd}}$. On the other hand, our results predict that the equivalent population of BCGs at $z=1$ will have typical Eddington ratios of $10^{-2}-10^{-4}{\rm L_{\rm Edd}}$. In analogy with black-hole binaries, such a strong Eddington ratio change corresponds to a state transition between the canonical ``low/hard'' state and extremely radiatively inefficient ``quiescent" state. 

Black-hole binaries are important tools in X-ray astronomy, often referred to as X-ray binaries and when radio loud, microquasars. In particular, observations show that these objects transit between various accretion states on time scales of days/years, and such short time scales allow detailed studies of how black holes evolve with time through monitoring campaigns \citep[see reviews by][]{McC2005,Rem200644}. Black-hole binary systems appear to typically transit between four major states (excluding the intermediate states): a very high state also known as the ``steep power-law'' state which is often accompanied by quasi-periodic oscillations (QPOs), a thermal state also known as the ``high/soft'' state characterized by an efficiently accreting black hole ($\apgt10^{-2}{\rm L_{\rm Edd}}$) in which the jet is quenched, a hard state known as the ``low/hard'' state where the black hole is accreting at low rates ($\aplt10^{-2}{\rm L_{\rm Edd}}$) and driving a powerful jet, and finally the quiescent state where a black hole is acrreting at rates orders of magnitude bellow the above rates. The luminosities associated with the quiescent state of black-hole binaries typically range between $L_{\rm X-ray}=10^{30.5-33.5}\ergps$ \citep{Rem200644}, equivalent to Eddington ratios in the range of $10^{-8}-10^{-5}{\rm L_{\rm Edd}}$ for stellar mass black holes. Some studies have suggested that the quiescent state is simply an extension of the low/hard state with even lower accretion rates. \citet{Gal2006370} showed that the fundamental plane of black-hole activity, an established relation between the mass, X-ray and radio luminosity of a black hole in the low/hard state, extends down to quiescence for black-hole binaries. Advection-dominated accretion flows (ADAFs) which have been able to explain the behaviour of black holes accreting at rates below a few per cent of Eddington have also been able to explain the behavior of quiescent black holes \citep[e.g.][]{Nar200851,Nar1996457,Nar1997482}, including Sgr A$^\star$, the quiescent black hole at the center of our galaxy that is accreting at \.{M}$=10^{-6}$\.{M}$_{\rm Edd}$ \citep{Yua2003598}.

Based on our knowledge of state transitions in black-hole binaries and the strong evolution seen in terms of Eddington ratios in our BCGs (Fig. \ref{fig5_8}), we propose that BCG-hosted black holes with X-ray cavities have migrated from the canonical ``low/hard'' state to the ``quiescent" state over the last 5 Gyrs. This would explain why the AGN in BCGs appear to be so radiatively inefficient at the present day. Quiescent states in black-hole binaries are long-lived; the objects spend most of their time in this state \citep[][90 per cent]{Don2007,Nar200851}. If BCG-hosted black holes with X-ray cavities are transiting into this state, then the black holes in these BCGs could provide a long-lived solution to the cooling-flow problem if jets are maintained throughout this state \citep[see also ][]{Dun2010404}.

The transition between the high/soft state to the low/hard state in black-hole binaries, as opposed to the transition between the low/hard state and the high/soft state, typically occurs at lower accretion rates, $0.02-0.05{\rm L_{\rm Edd}}$ as opposed to $0.1-0.2{\rm L_{\rm Edd}}$ \citep[][]{Miy1995442,Mac2003338,Zdz2004351,Don2007}. At least in terms of BCGs in strong radio-mode feedback with clear X-ray cavities, our results suggest that a significant population may host powerful quasars at their centres at $z\approx1$, and some of the currently known $z\approx1$ cool-core clusters fit in this category. These are accreting at rates near $10^{-2}{\rm L_{\rm Edd}}$, which places them at the limit of the transition state between the high/soft and low/hard state analogous to black-hole binaries. This would have occurred in the last 7 to 8 Gyrs, while black-hole binaries take roughly 1 year to transit between these states \citep[e.g. G1655-10;][]{McC2005,Rem200644}. Scaling the black-hole binary time scale from a $1-5{\rm M_\odot}$ to a $10^{10}{\rm M_\odot}$ black hole fits well with the observed time scale for BCGs. 

We further note that some black-hole binaries in quiescent states exhibit chaotic variations in their luminosities (i.e. ``flaring''), while others remain stable for decades \citep[e.g. A06200-00,][]{Gal2007670}. V404 Cyg is one example and consists of a black-hole binary system that has a typical X-ray luminosity of $10^{33}\ergps$ in quiescence but its luminosity is known to vary by a factor of a few within hours \citep{Wag1994429,Kon2002570,Hyn2003345}. Another example is Sgr A$^\star$, where the flare fluxes can be a factor of a few to ten times those of the quiescent state and appear on minute time scales \citep{Bag2001413,Por2003407,Por2008488}. The chaotic behaviour seen in some quiescent black-hole binaries could be analogous to what is seen in Perseus, where the luminosity of the central AGN has varied by a factor of a few in the past decade \citep[see different flux values by][]{All2001322,Eva2006642,Chu2003590}. Careful X-ray monitoring of this source would be needed to disentangle the variations in the nucleus emission.

\subsection{Implications for cluster surveys}

Finally, we mention the implications our results have concerning cluster surveys and the identification of high-redshift cool-core clusters. Based on the results for PKS 1229-021, a $z\approx1$ quasar embedded in a cool-core cluster of galaxies, \citet{Rus2012422} suggested that high-redshift cluster surveys could be missing a significant fraction of cool-core clusters if the central BCG has a quasar-like luminosity \citep[see also][]{All201149}. 

Our results suggest that the currently known population of BCGs with X-ray cavities may be evolving with cosmic time, such that many may have $10^{45-46}\ergps$ $2-10\keV$ luminosities at $z=1$. These are quasar-like luminosities, implying that a significant fraction of BCG-hosted AGN at $z=1$ could harbour a quasar in their centres and therefore outshine the majority of the cluster X-ray emission. Our results therefore suggest that when cluster surveys are conducted at high redshift based on moderate resolution X-ray imaging (e.g. $ROSAT$ Position Sensitive Proportional Counters), a significant population of cool-core clusters might be missed since they would appear as point-like AGN and not as extended X-ray objects. This identification bias could explain, at least in part, the apparent lack of $very$ strong cool-core clusters at high redshift \citep[$z>0.5$;][]{Vik2007,San2008483,San2010521,All201149}. 3C 186 \citep{Sie2010722} and PKS 1229-021 are two such examples of bright quasars embedded in cool-core clusters at $z\approx1$ where the X-ray emission from the BCG dominates over the cluster emission.

Note that, our interpretation assumes that AGN duty cycles remain high at high redshift. We base this assumption on the middle panel of Fig. \ref{fig5_10}, which shows that the majority of our $z=0.5$ BCGs have a bright, radiatively efficient AGN (60 per cent). If this high fraction persists up to $z=1$, then we can assume that the majority of high-redshift BCGs in cool core clusters will have a bright central AGN. In other words, high-redshift BCGs in cool core clusters would spend most of their time X-ray bright, so AGN duty cycles would be high, at least in terms of the radiative efficiency. 

At optical wavelengths, BCGs often exhibit Seyfert-like line emission without being X-ray point sources \citep[see][for a detailed look at optical properties of BCGs]{Cra1999306}. The X-ray emission associated with the Seyfert-type BCG could therefore appear point-like in cluster surveys, yet originate from the hot thermal gas of the intracluster medium. Existing optical classifications of line-emitting BCGs can therefore cause cluster surveys to miss cool-core clusters, especially at high redshift. BCGs in cool core clusters may also be becoming increasingly blue at high redshift, complicating even more the search for such objects if the searches are based on the red sequence \citep[see from a theoretical point of view,][]{DeL2007375}. 

In summary, the existence of a significant population of high redshift clusters with quasars at their centres complicates the search for such objects, at least for many of the techniques currently used to find them. The current population of high redshift cool core clusters may therefore be undersampled and the lack of strong cool core clusters observed from surveys \citep[][]{Vik2007,San2008483,San2010521,All201149} may simply be due, at least in part, to selection effects.

\section{Concluding remarks}

We have investigated the evolution of the nuclear $2-10\keV$ luminosity in 32 BCGs that lie in highly X-ray luminous clusters of galaxies ($L_{\rm X-ray~}{\rm (0.1-2.4\keV)}>3\times10^{44}\ergps$) and that have known X-ray cavities. We stress that we are only considering BCGs in which strong radio-mode feedback is taking place in the form of AGN-driven X-ray cavities, and therefore only considering a sub-population of BCGs. We further stress that the detectability of X-ray cavities most likely depends on the redshift of the sources. To limit the possible selection effects, we have applied stringent selection criteria to our sample, and only consider highly X-ray luminous clusters where X-ray cavities can be more easily identified. 

Applying these criteria, we find evidence for evolution such that the average nuclear luminosity in these BCGs has increased by at least a factor of 10 from $0<z<0.3$ to $0.3<z<0.6$. If we further limit the sample to clusters that have similar cooling luminosities accross all redshifts (26 in total), we find that the scatter decreases and the increase in brightest with redshift remains the same. 

Mindful of potential biases in our sample, we propose that the central AGN of currently known X-ray luminous clusters with X-ray cavities are steadily becoming fainter over time, or more likely, that the fraction of BCGs with radiatively efficient nuclei is decreasing with time from roughly 60 per cent to 30 per cent over the last 5 Gyrs. In analogy with black-hole binaries and based on the observed change in the Eddington ratios of our sources, we further propose that the evolving AGN population in BCGs may be transiting from a canonical low/hard state, analogous to that of X-ray binaries, to a quiescent state over the last 5 Gyrs. Our results also suggest that a significant fraction of BCGs in $z\approx1$ clusters may host quasars at their centres, potentially complicating the search for such clusters at high redshift. 

New surveys based on the Sunyaev-Zel'dovich effect which has no surface brightness dimming (SPT, $Planck$) will soon be online and will provide a wealth of new data in the intermediate to high redshift Universe. These, coupled with follow-up $Chandra$ observations may significantly enhance the current sample of clusters with known X-ray cavities, thus providing more insight into the potential evolution observed in this study. Extending the analysis to the general population of BCGs, and examining the evolution of the X-ray nuclear luminosity across the entire population may also provide a better understanding of the evolution observed here.

\section*{Acknowledgments}

JHL recognizes the support given by the Cambridge Trusts, Natural Sciences and Engineering Research Council of Canada (NSERC), as well as the Fonds Quebecois de la Recherche sur la Nature et les Technologies (FQRNT). ACF thanks the Royal Society. We thank Helen Russell for providing the cooling time profile of H1821+643. SWA acknowledges support from the U.S. Department of Energy under contract number DE-AC02-76SF00515. JHL is also supported by NASA through the Einstein Fellowship Program, grant number PF2-130094.

\label{lastpage}
\bibliographystyle{mn2e}
\bibliography{bibli}

\appendix
\section{Detailed X-ray spectral modelling}
In this section, we present the different nuclear models adopted for each of the 11 objects in our sample that show evidence of an X-ray point source in their $Chandra$ images. These exclude Perseus and H1821+643 since we use the nuclear fluxed quoted the literature for these sources. For each of the 11 objects, we extract a $0.5-7\keV$ spectrum within a 1$''$ circular region centred on the X-ray point source. We then take a surrounding annulus located within $2''$ and 3$''$ as the background. All of our fits include Galactic absorption which we keep frozen at the \citet{Kal2005440} value. We use C-statistics to account for the low number of counts.

RXC J1524.3$-$3154 - The background-subtracted nuclear spectrum of this source is very noisy, and requires several parameters to be frozen in order to constrain the fit. If we fit a simple absorbed (Galactic) {\sc mekal} model to the spectrum, and freeze the abundance at the value obtained from a {\sc mekal} model applied to the $2''-3''$ surrounding annulus, we obtain an abnormally large temperature ($13^{+24}_{-7}\keV$). We therefore investigate other models that can explain the observed spectrum, and consider both a power-law model without (Model I) and with (Model II) internal absorption. The results are shown in Table \ref{tab5_5}. 

RXC J1558.3$-$1410 - The background-subtracted nuclear spectrum of this source cannot be fitted with an absorbed (Galactic) {\sc mekal} model. The fit is largely unconstrained, even if the abundance is kept frozen. If we fit a simple absorbed (Galactic) power-law, the power-law index tends towards a negative value. This is due to two components being present in the spectrum, one at soft X-rays ($\approx1\keV$) and another at hard X-rays ($\approx4\keV$). Even an absorbed (Galactic + internal) power-law model cannot fit the soft X-rays properly, and the internal absorption tends towards a null value. Complex absorbers do not provide good fits either. However, fitting a double power-law, where one of the power-laws is internally absorbed ({\sc zphabs}), provides a good fit if we keep the photon index of the second, internally absorbed power-law frozen at a value of 1.9 (Model I). We also investigate the possibility that the emission seen at soft X-rays is of thermal origin. If we fit an absorbed (Galactic) {\sc mekal + power-law} model to the data and let the temperature, abundance, normalization parameters, as well as photon index value vary, the fit is largely unconstrained. To provide a better constraint on the model, we adopt the method used in HL2011 to derive the non-thermal flux. In this case, the surrounding $2''-3''$ annulus is not used as the background. Instead, it is used to estimate the properties of the cluster thermal component which are then extrapolated to the inner 1$''$ circular region. The background is chosen as a region located far from cluster emission. For the surrounding $2''-3''$ annulus, we fit an absorbed (Galactic) {\sc mekal} model to the data and find the best-fitting temperature, abundance and normalization parameters. The extracted temperature is then extrapolated down to the $1''$ circular region assuming that $T=ar^b$ where $b\approx0.3$ \citep{Voi2004347}. The abundance is not expected to vary significantly from $r=1''$ to $r=2''-3''$. Using this abundance and extrapolated temperature, we first fit an absorbed (Galactic) {\sc mekal + pow} model to the nuclear spectrum. We let both normalization parameters and photon index free to vary. We also add the constraint that the normalization of the thermal component is not allowed to be less than the normalization obtained in the fit for the annulus ($r=2''-3''$), scaled for the same pixel number, since the density is expected to increase with decreasing radius. In this case, the power-law index tends towards a negative value. We therefore add an internal absorption at the redshift of the source, while keeping the power-law index frozen at a value of 1.9 to help constrain the fit (Model II).

4C+55.16 - Fitting a simple absorbed (Galactic) {\sc mekal} model to the data provides a good fit, but the resulting temperature is abnormally high ($\approx11\keV$). This might indicate shock heating in the vicinity of the central regions, but we also investigate the possibility that the emission is of non-thermal origin since the temperature jump is significant. If we fit an absorbed power-law model to the data, where the absorption accounts for both Galactic and internal absorption at the redshift of the source, we find that the internal absorption tends towards a null value. A simple power-law, where the absorption accounts only for Galactic absorption provides a good fit and we show the results in Table \ref{tab5_5} (Model I). Fitting more complex absorbers ({\sc zpcfabs, pwab}) does not provide a good fit to the data.

MACS J0547.0$-$3904 - The background-subtracted nuclear spectrum is dominated by emission seen around $1\keV$. We begin by fitting a simple absorbed power-law to the background-subtracted spectrum, where the absorption accounts only for Galactic absorption (Model I). The resulting parameters and unabsorbed nuclear $2-10\keV$ luminosity are shown in Table \ref{tab5_5}. Although the model provides a reasonable fit, it underestimates the hard X-rays ($>3\keV$). Adding an additional absorption at the redshift of the source provides a better fit, but the resulting power-law index is $\approx5$, which is unlikely. Instead, we interpret the emission seen around $1\keV$ as thermal emission. Fitting a simple {\sc mekal} model to the data does not provide a good fit, and significantly underestimates the hard X-rays ($>2\keV$). There is clearly a non-thermal contribution that dominates beyond 2$\keV$. If we fit an absorbed (Galactic) {\sc mekal + power-law} model to the data, and let the temperature, abundance, normalization parameters, as well as photon index value vary, the fit is largely unconstrained. We therefore adopt the same technique as in RXC J1558.3$-$1410 to fit the data. In this case, the surrounding $2''-3''$ annulus is used to constrain the parameters of the plasma model, and the background is taken as a region located far from any cluster emission. We then fit a {\sc mekal} and internally absorbed power-law model to the data (Model II). Note that, other more complex absorbers, such as partial covering absorption models ({\sc zpcfabs, pwab}) do not provide good fits.

MACS J1931.8$-$2634 - The background-subtracted nuclear spectrum is clearly dominated by non-thermal emission; an absorbed (Galactic) {\sc mekal} model provides a very poor fit even if we limit the fitting range to the $0.5-2\keV$ energy band. We therefore concentrate on fitting different non-thermal models to the spectrum. A simple absorbed (Galactic) power-law does not provide a good fit. However, a power-law with additional internal absorption at the redshift of the source is able to reproduce the observed spectrum (see Table \ref{tab5_5}, Model I). We also try fitting a double power-law model to the data, but the model is highly unconstrained, and more complex absorbers ({\sc zpcfabs, pwab}) are not able to reproduce the spectrum. The luminosity we find agree well in the results of \citep{Ehl2011411}.

MACS J0947.2+7623 - Fitting an absorbed (Galactic + internal {\sc zphabs}) power-law to the background-subtracted nuclear spectrum does not provide a good fit at energies below $1\keV$. A different absorbing component at the redshift of the source is needed to explain the emission. \citet{Cav2011732} modelled the nuclear emission using a partially covered absorber ({\sc zpcfabs}), as well as a power-law and 2 gaussian lines to account for features seen around 1.8 and 3$\keV$. However, by using their model, we were not able to constrain the parameters, especially those for the additional gaussian lines. Their model essentially requires several parameters to be frozen in order to constrain the fit. \citet{Cav2011732} also fitted a distribution of partially covered absorbers with the {\sc pwab} model, but this model is also unable to converge to a solution. We therefore choose to calculate the flux of the non-thermal component in this source by fitting an absorbed (Galactic + {\sc zpcfabs}) power-law model to the emission (Model I). We also include a model with an additional gaussian line around $3\keV$, but keep the width frozen at 100$\eV$ to help constrain the fit (Model II).

MACS J2046.0$-$3430 - Fitting an absorbed (Galactic) {\sc mekal} model to the background-subtracted spectrum does not provide a good fit, and underestimates the emission beyond $2\keV$. This suggest that there is a significant non-thermal contribution to the spectrum. If we fit a simple power-law model to the spectrum with Galactic absorption, the resulting index is unnaturally large ($\Gamma>4$). Including internal absorption at the redshift of the source does not improve the fit, and the internal absorption converges towards a null value. Instead, we consider the possibility that there is a contribution of both thermal and non-thermal emission to the nuclear spectrum, and apply the same method as for RXC J1558.3$-$1410 to help constrain the fit. In this case, the surrounding $2''-3''$ annulus is used to constrain the thermal plasma parameters within the $1''$ circular region, and the background is taken as a region located far from any cluster emission. We consider power-law models without (Model I) and with (Model II) internal absorption at the redshift of the source, but keep the index frozen ($\Gamma=1.9$) to help constrain the fit.

MACS J0913.7+4056 - The background-subtracted nuclear spectrum, where the background is taken as the surrounding $2''-3''$ annulus, shows three distinct components: one at soft X-rays ($1\keV$) and two at hard X-rays (a bump around $3\keV$ and an emission line at $4.6\keV$ with a $\approx$60$\eV$ width). The location of the emission line coincides with the redshifted 6.4$\keV$ Fe K emission complex. A simple absorbed (Galactic) power-law with an additional emission line does not provide a good fit at soft X-rays, even if we add an internal absorption at the redshift the source. However, the emission seen around $1\keV$ can be modelled by adding a complex absorber at the redshift of the source ({\sc zpcfabs}). The fit is constrained if we keep both the energy (4.6$\keV$) and width (60$\eV$) of the line frozen. The resulting values are shown in Table \ref{tab5_5} (Model I). We also consider the possibility that the emission seen around $1\keV$ is of thermal origin. In this case, we model the background-subtracted nuclear spectrum as an absorbed (Galactic) {\sc mekal} and power-law with emission line model, where we keep the emission line energy and width frozen. We also keep the abundance of the thermal component frozen at the best-fitting value for the surrounding $2''-3''$ annulus. The fit is constrained, but the best-fitting power-law index is negative. If we add an internal absorption at the redshift of the source ({\sc zphabs}), the model is not able to fit properly the emission seen at soft X-rays, and the upper limit of the temperature is not constrained. Freezing the temperature at a similar value derived from a {\sc mekal} model fitted to the surrounding $2''-3''$ annulus does not provide a good fit. It is therefore more likely that the emission seen around $1\keV$ is of non-thermal origin and we model the $0.5-7\keV$ emission as a double power-law, where one of the power-laws is also affected by internal absorption at the redshift of the source (Model II). To help constrain the fit, we keep the energy and width of the emission line frozen, and first fit the $0.5-2\keV$ energy range to constrain the power-law index of the first power-law (at soft X-rays). This yields a value of $1.6\pm0.5$. We then fit the complex model to the entire $0.5-7\keV$ range and keep the index of the first power-law frozen at 1.6. The nuclear spectrum of this source has been looked at in detail by authors \citep{Iwa2001321,Pio2007473,Vig2011416}.

MACS J1411.3+5212 - The nuclear spectrum of this source is similar to the one in MACS J0913.7+4056. There are two distinct components: one at soft X-rays ($\approx1\keV$) and one at hard X-rays ($\approx3\keV$). A simple absorbed power-law cannot account for both components and a more complex model is needed to explain the emission. We begin by fitting an absorbed (Galactic) power-law with a partial covering absorber at the redshift of the source ({\sc zpcfabs}) and show the results in Table \ref{tab5_5} (Model I). The spectrum can also be fitted with a double power-law model, where one of the power-laws is affected by additional internal absorption at the redshift of the source (Model II). However, we also consider the possibility that the emission seen at soft X-rays is of thermal emission. We keep the abundance frozen at the best-fitting value of a {\sc mekal} model applied to the surrounding $2''-3''$ annulus (0.62Z$_{\rm \odot}$) and let the temperature, normalization parameters, internal absorbing column density and power-law index free to vary (Model III). 

MACS J1423.8+2404 - We begin by fitting a {\sc mekal} model to the background-subtracted nuclear spectrum. Although the fit converges, it slightly underestimates the emission seen at hard Xrays ($>3\keV$). Instead, we try fitting an absorbed (Galactic) power-law to the spectrum (Model I), which provides a better fit at hard X-rays. We also consider the possibility that there is both a thermal and non-thermal contribution to the spectrum, and apply the same method used for RXC J1558.3$-$1410 to help constrain the fit. This method consists of using the surrounding $2''-3''$ annulus as a proxy for the thermal plasma parameters within the 1$''$ circular region. Applying this method, we fit a {\sc mekal} an internally absorbed power-law model to the nuclear spectrum. The background is taken as a background far from any cluster emission. For the fit to converge, we also require the power-law index to be frozen. In this case, the internal absorption converges towards a null value. We therefore choose to fit the nuclear spectrum with a simple {\sc mekal} and absorbed power-law model (Model II). 

\begin{figure}
\centering
\begin{minipage}[c]{0.99\linewidth}
\centering \includegraphics[width=\linewidth]{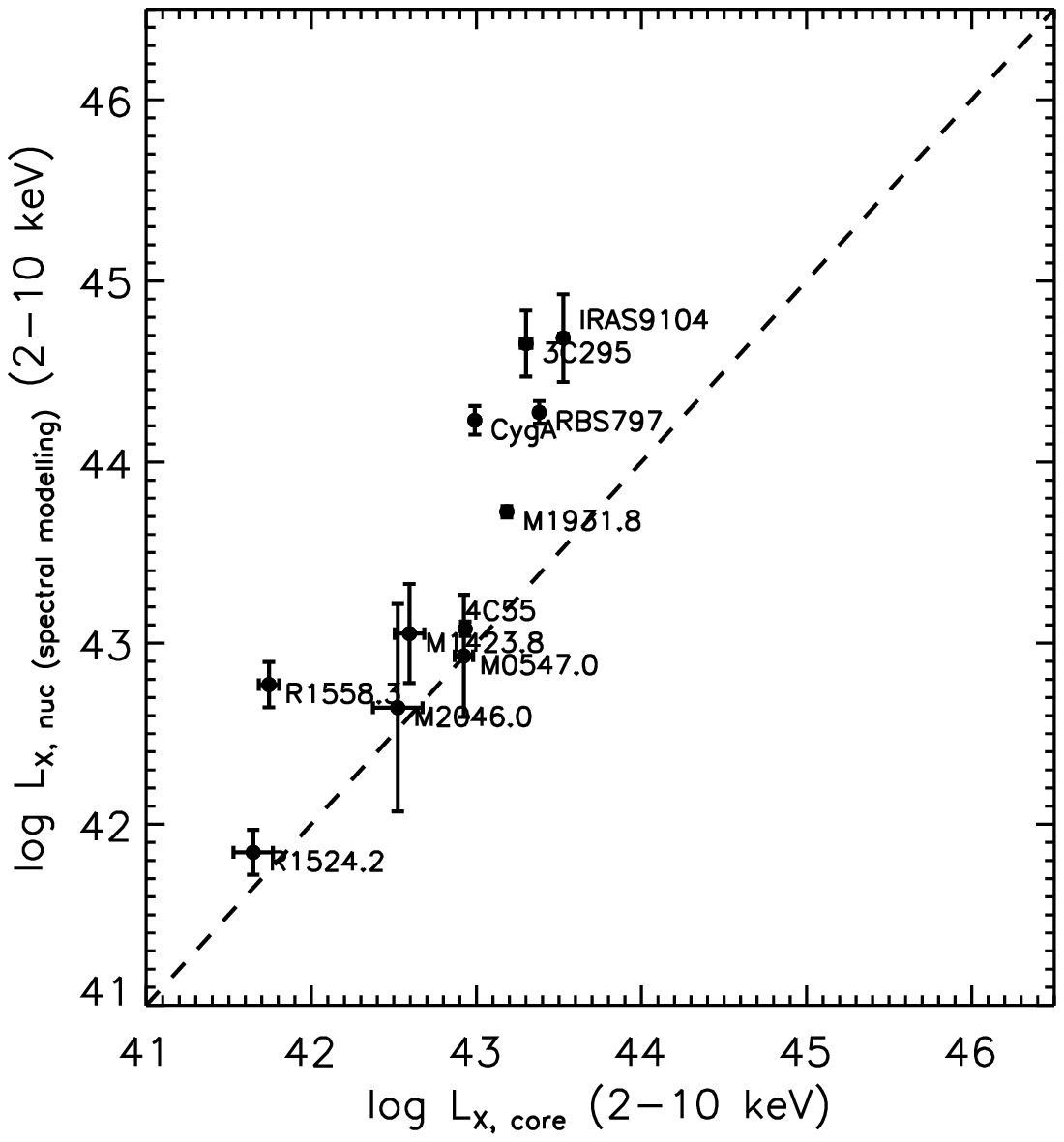}
\end{minipage}
\caption[]{Logarithm of the $2-10\keV$ nuclear luminosity derived from spectral modelling as a function of the luminosity derived from the core method (see Section 4.1). The latter is derived from an observed flux, and therefore does not correct for internal absorption. We only show the 11 sources where we performed a spectral modelling analysis. Equality is illustrated with the dashed line. }
\label{figApp1}
\end{figure}

Cygnus A - The $Chandra$ nuclear spectrum of this source shows clearly the presence of a non-thermal component at hard X-rays which can be modelled as an absorbed (intrinsic) power-law with an additional gaussian line. However, there is also the presence of a faint soft X-ray component that we first model as a power-law. Here, we keep the gaussian line energy and width fozen at a value of 6.1$\keV$ and of width 0.05$\keV$ to help constrain the fit (Model I). We also consider the possibility that the emission seen is of thermal nature, and apply the same method as for RXC J1558.3$-$1410 to help constrain the fit. In this case, the gaussian line energy and width are also kept fozen at a value of 6.1$\keV$ and of width 0.05$\keV$ to help constrain the fit (Model II).

Note that in HL2011, we performed a spectral analysis for many of the sources with no detectable X-ray nucleus such as Abell 1835 in an attempt to further constrain the upper limit of the non-thermal fluxes. Although we were able to obtain a rough estimate of the non-thermal fluxes for these objects, by applying the same method as for  RXC J1558.3$-$1410 (i.e. the method where we use the surrounding annulus to estimate the {\sc mekal} plasma parameters of the inner 1$''$ circular region), we stressed that most of the parameters in the analysis had to be frozen in order to constrain the fit. It was also not clear if the addition of the power-law component for these objects actually improved the fit, and we proposed that these apparent detections could be simply due to the manifestation of cooler thermal components which is often seen in the central regions of cool-core clusters. We therefore assume that it is reasonable to treat all the derived luminosities for the 19 objects in our sample with no detectable X-ray nucleus as upper limits. 

In Fig. \ref{figApp1}, we compare the spectral modelling results to those obtained in Section 4.1 from the core emission. Since the spectral analysis corrects for internal absorption at the redshift of the source (if present), the derived values should be equal or larger than those derived from the core emission. This is what is seen in Fig. \ref{figApp1}.

\end{document}